\documentclass[fleqn,usenatbib,useAMS]{mnras}

\usepackage{multirow}
\usepackage{amssymb, amsmath}

\usepackage{graphicx}	
\usepackage{amsmath}	
\usepackage{amssymb}	
\usepackage{multicol}        
\usepackage{bm}		
\usepackage{pdflscape}	
\usepackage{orcidlink}
\usepackage{newtxtext,newtxmath}

\DeclareRobustCommand{\VAN}[3]{#2}
\let\VANthebibliography\thebibliography
\def\thebibliography{\DeclareRobustCommand{\VAN}[3]{##3}\VANthebibliography}

\usepackage[T1]{fontenc}
\usepackage{ae,aecompl}

\usepackage{newtxtext,newtxmath}

\newcommand{\GHz}{{\rm\,GHz}}
\newcommand{\Jy}{{\rm\,Jy}}
\newcommand{\alphasyn}{\alpha_{\rm syn}}
\newcommand{\HEALPix}{{\tt HEALPix}}
\newcommand{\Planck}{\textit{Planck}}
\newcommand{\mum}{\,$\mu{\rm m}$}
\newcommand{\Aame}{{A_{\rm AME}}}

\newcommand{\Same}{{S^{\rm AME}}}
\newcommand{\nuame}{{\nu_{\rm AME}}}
\newcommand{\Wame}{{W_{\rm AME}}}
\newcommand{\taud}{{\tau_{\rm353}}}
\newcommand{\betad}{{\beta_{\rm d}}}
\newcommand{\Td}{{T_{\rm d}}}
\newcommand{\EM}{{{\rm EM}}}

\newcommand{\Asyn}{{{A_{\rm1\,GHz}}}}
\newcommand{\chisqred}{{\chi^2_{\rm red}}}
\newcommand{\emmAME}{{\epsilon_{\rm AME}^{\rm 28.4\,GHz}}}
\newcommand{\nbeam}{{n_{\rm beam}}}
\newcommand{\lt}{<}
\newcommand{\AIC}{{\rm AIC}}
\newcommand{\BIC}{{\rm BIC}}

\newcommand{\Tff}{{T_{\rm ff}}}
\newcommand{\kb}{{k_{\rm B}}}
\newcommand{\h}{{\rm h}}

\title[AME in M31 with QUIJOTE-MFI]{QUIJOTE Scientific Results -- XVII. 
Studying the Anomalous Microwave Emission in the Andromeda Galaxy with QUIJOTE-MFI}
\author[Fern\'{a}ndez-Torreiro et al.]
{{M. Fern\'{a}ndez-Torreiro$^{\orcidlink{0000-0002-6805-9100}}$,$^{1,2}$\thanks{E-mail: mateo.fernandez@iac.es (MFT)}}
{R.~T. G\'{e}nova-Santos$^{\orcidlink{0000-0001-5479-0034}}$,$^{1,2}$}\thanks{E-mail: rgs@iac.es (RTGS)}
{J.~A. Rubi\~{n}o-Mart\'{\i}n$^{\orcidlink{0000-0001-5289-3021}}$,$^{1,2}$}
\newauthor
{C.~H. L\'{o}pez-Caraballo$^{\orcidlink{0000-0002-6439-5385}}$,$^{1,2}$}
{M.~W. Peel$^{\orcidlink{0000-0003-3412-2586}}$,$^{1,2,3}$}
{C.~Arce-Tord$^{\orcidlink{0000-0002-0176-4331}}$, $^{4,5}$}
{R. Rebolo$^{\orcidlink{0000-0003-3767-7085}}$,$^{1,2,6}$}
{E. Artal$^{\orcidlink{0000-0002-2569-1894}}$,$^{7}$}
\newauthor
{M. Ashdown$^{\orcidlink{0000-0003-2044-7523}}$,$^{8,9}$}
{R.~B. Barreiro$^{\orcidlink{0000-0002-6139-4272}}$,$^{10}$}
{F.~J. Casas$^{\orcidlink{0000-0002-2217-5843}}$,$^{10}$}
{E. de la Hoz$^{\orcidlink{0000-0002-5066-816X}}$,$^{10,11,12}$}
{F. Guidi$^{\orcidlink{0000-0001-7593-3962}}$,$^{13}$}
\newauthor
{D. Herranz$^{\orcidlink{0000-0003-4540-1417}}$,$^{10}$}
{R. Hoyland$^{\orcidlink{0000-0001-5346-0519}}$,$^{1,2}$}
{A. Lasenby$^{\orcidlink{0000-0002-8208-6332}}$,$^{8,9}$}
{E. Mart\'{i}nez-Gonzalez$^{\orcidlink{0000-0002-0179-8590}}$,$^{10}$}
L. Piccirillo,$^{14}$
\newauthor
{F. Poidevin$^{\orcidlink{0000-0002-5391-5568}}$,$^{1,2}$}
{B. Ruiz-Granados$^{\orcidlink{0000-0003-3229-2725}}$,$^{1,2,15}$}
D. Tramonte,$^{16,1,2}$
F. Vansyngel,$^{1,2}$
{P. Vielva$^{\orcidlink{0000-0003-0051-272X}}$,$^{10}$}
\newauthor
{R.~A. Watson$^{\orcidlink{0000-0002-5873-0124}}$.$^{14}$}
\\
$^{1}$Instituto de Astrof\'{\i}sica de Canarias, E-38200 La Laguna, Tenerife, Spain\\
$^{2}$Departamento de Astrof\'{\i}sica, Universidad de La Laguna,
E-38206 La Laguna, Tenerife, Spain\\
$^{3}$Imperial College London, Blackett Lab, Prince Consort Road, London SW7 2AZ, UK \\
$^{4}$Departamento de Astronom\'{\i}a, Universidad de Chile, Camino El Observatorio 1515, Las Condes, Santiago, Chile \\
$^{5}$Millennium Nucleus on Young Exoplanets and their Moons (YEMS), Chile.\\
$^{6}$Consejo Superior de Investigaciones Cient\'{\i}ficas, E-28006
Madrid, Spain\\$^{7}$Universidad de Cantabria, Departamento de Ingeniería de Comunicaciones, Edificio Ingenieria de Telecomunicación, Plaza de la Ciencia nº 1, 39005 Santander, Spain\\
$^{8}$Astrophysics Group, Cavendish Laboratory, University of Cambridge, 
J J Thomson Avenue, Cambridge CB3 0HE, UK\\
$^{9}$Kavli Institute for Cosmology, University of Cambridge, Madingley Road, Cambridge CB3 0HA, UK\\
$^{10}$Instituto de F\'{\i}sica de Cantabria (IFCA), CSIC-Univ. de Cantabria, Avda. los
Castros, s/n, E-39005 Santander, Spain\\
$^{11}$Departamento de F\'{\i}sica Moderna, Universidad de Cantabria,
Avda. de los Castros s/n, 39005 Santander, Spain\\
$^{12}$CNRS-UCB International Research Laboratory, Centre Pierre Binétruy, IRL2007, CPB-IN2P3, Berkeley, CA 94720, USA\\
$^{13}$Institut d'Astrophysique de Paris, UMR 7095, CNRS \& Sorbonne Universit\'e, 98 bis boulevard Arago, 75014 Paris, France\\
$^{14}$Jodrell Bank Centre for Astrophysics, Alan Turing Building, Department of Physics \& Astronomy, School of Natural Sciences, The University of Manchester, \\ Oxford Road, Manchester, M13 9PL, U.K. \\
$^{15}$Departamento de F\'{\i}sica. Facultad de Ciencias. Universidad de C\'ordoba. Campus de Rabanales, Edif. C2. Planta Baja.  E-14071 C\'ordoba, Spain.\\
$^{16}$Department of Physics, Xi'an Jiaotong-Liverpool University, 111 Ren'ai Road, \\    \quad Suzhou Dushu Lake Science and Education Innovation District, Suzhou Industrial Park, Suzhou 215123, P.R. China.}

\date{Accepted 2023 October 10. Received 2023 October 10; in original form 2023 May 12}

\pubyear{2023}

\begin{document}
\label{firstpage}
\pagerange{\pageref{firstpage}--\pageref{lastpage}}
\maketitle
\defcitealias{battistelli2019}{B19}

\begin{abstract}
The Andromeda Galaxy (M31) is the Local Group galaxy that is most 
similar to the Milky Way (MW). The
similarities between the two galaxies make M31 useful for 
studying integrated properties common to spiral galaxies.
We use the data from the recent QUIJOTE-MFI Wide Survey, 
together with new raster
observations focused on M31, to study its integrated
emission. The addition of raster data improves the
sensitivity of QUIJOTE-MFI maps by almost a factor 3.
Our main interest is to confirm if anomalous
microwave emission (AME) is present in M31, as previous
studies have suggested.
To do so, we built the integrated spectral energy 
distribution of M31 between 0.408 and 3000\GHz{}.
We then performed a component separation analysis
taking into account synchrotron, free-free, AME and
thermal dust components.
AME in M31 is modelled as a log-normal distribution with
maximum amplitude, $\Aame$, equal to $1.03\pm0.32$\,Jy.
It peaks at $\nuame=17.2\pm3.2\GHz{}$ with a width of
$\Wame=0.58\pm0.16$. Both the Akaike and Bayesian 
Information Criteria find the model without AME to be less
than 1\% as probable as the one taking AME into consideration.
We find that the AME emissivity per 100$\,\mu$m 
intensity in M31 is 
$\emmAME=9.6\pm3.1\,\mathrm{\mu}$K/(MJy/sr), similar
to that of the MW. We also provide the first 
upper limits for the AME polarization fraction in an
extragalactic object.
M31 remains the only galaxy where an AME measurement
has been made of its integrated spectrum.
\end{abstract}




\section{Introduction}
\label{section:introduction}
The Andromeda Galaxy or Messier 31 (M31) is the largest
galaxy in the Local Group \citep{M31_MW_radii} and the most similar
to the Milky Way (MW), both being spiral galaxies. It 
has been observed and studied in detail throughout modern
astronomical history owing to its large angular size,\footnote{M31's 
isophotal major radius is 91.5\arcmin
\citep{m31_isophot_radius} with an axial ratio of $\approx0.7$.  
The Herschel 
Exploitation of Local Galaxy Andromeda 
\citep[HELGA,][]{herschelM31} program went beyond M31 itself
and also studied its surroundings, producing maps of
$\sim5.5\degr\times2.5\degr$.} over 5\,deg$^2$.
These studies span the whole
frequency domain: a review of high resolution radio
measurements is available in \cite{berkhuijsen2003}.
\cite{fatigoni2021} also studied M31 at high angular
resolution, using data taken in the C-band of the Sardinia Radio
Telescope (hereafter, SRT). Together with
the previously mentioned ancillary data, thermal and
non-thermal emission components could be spatially
disentangled, and the star formation rate was estimated.

Anomalous microwave emission (hereafter, AME) is a 
prominent mechanism of Galactic emission in the 
frequency range 10--60 GHz. Known for more than
25 years now \citep{kogut1996, leitch1997}, the 
physical process responsible for AME is not clear yet.
The dominant hypothesis states that spinning dust emission
is responsible for AME \citep{draine1998a, draine1998b},
owing partly to the strong spatial 
correlation between AME and thermal dust (300--3000\GHz) 
emission \citep[e.g.,][]{AMEwidesurvey,ameplanewidesurvey}.
However, the exact carrier responsible for AME is still
a mystery. Neither are its polarization properties
 well understood  yet, although observations both on compact regions at angular scales of $1\degr$ or below \citep[e.g.,][]{caraballo2011, dickinson2011, Perseus, W44} and on large angular scales \citep{macellari2011,herman2022} indicate that its polarization fraction should be below 5\%. See \cite{amereview} for a
detailed review on AME.

While there have been many detections of AME in our Galaxy 
\citep[e.g.,][]{watson2005, planck2015galacticcloudsAME, Taurus,
LambdaOrionis, ameplanewidesurvey} 
extragalactic detections of AME are scarce \citep{murphy2010, 
murphy2018}. \cite{peel2011} set upper limits
on the AME from three different bright galaxies using
the WMAP 7 yr Point Source Catalogue and the \Planck{}
Early Release Compact Source Catalogue (ERCSC).
Using observations from these two satellites, \cite{planckM31}
claimed a 2.3$\sigma$ significance detection of AME in 
M31. \cite{tibbs2018M33} also analysed \Planck{} data 
for M33 but found no evidence of an AME component on
its integrated spectrum.
\cite{SRTgalaxies2022} studied data from K-band 
detectors at SRT for four nearby spiral galaxies but was
unable to detect AME in any of them, which instead places constraints on its
upper levels. However, AME has been detected in 
resolved extragalactic regions, such as a star-forming
region in NGC 6946 \citep{murphy2010, hensley2015murphy2010}
and a compact radio source associated with NGC 4725 
\citep{murphy2018}. It appears that AME is much easier to
detect in individual, well isolated regions, where the background
emission can be easily subtracted. Generally, these regions
also show high emission from thermal dust and high star
formation ratios, together with low synchrotron signal.
It is difficult to find a galaxy whose integrated spectrum
shows this combination of properties, as normally many
strong components contribute to the spectrum. 

Based on new data at 6.6 GHz from the SRT, which are key to 
determining the level of free--free emission, \cite{battistelli2019}
(hereafter, \citetalias{battistelli2019}) claimed a 
high-significance detection of AME in M31. These SRT data
were smoothed to 1 degree resolution and studied together
with data from WMAP, \Planck{} and \textit{Herschel} 
\citep{herschelM31} satellites. Few datasets
between 2 and 20\GHz{} covering M31 are currently available. This causes 
synchrotron, free--free and AME models to be highly 
degenerate unless strong priors are placed on their 
parameters. However, the combination of SRT data
with lower frequency (below 2\GHz{}) surveys
permitted a good determination of the synchrotron and 
free--free levels. Once synchrotron and free--free have been
properly defined, a good determination of AME is 
straightforward. The significance of the AME
detection increased from 2.3$\sigma$ in \cite{planckM31} to 
over 8$\sigma$ in \citetalias{battistelli2019} after including 
these SRT data. 

\cite{cbassm31} performed an independent analysis
of M31 after adding data from the C-BASS experiment at 4.76\GHz{}.
This data addition should improve the disentanglement of synchrotron
and free--free, similar to what happened with SRT data.
Using the same apertures as \cite{planckM31}, they claimed a 
3.0$\sigma$ detection of AME, although with much lower amplitude
than that expected from both \cite{planckM31} and 
\citetalias{battistelli2019}.
\cite{planckM31} and \cite{cbassm31} obtained $S_{\rm AME}^{\rm30\GHz{}}=0.7\pm0.3$\Jy{} and $0.27\pm0.09$\Jy{}, respectively, while using a smaller aperture \citetalias{battistelli2019} obtained $S_{\rm 
AME}^{\rm25\GHz{}}=1.45\pm0.15$\Jy{}. As a consequence of their low AME amplutide \cite{cbassm31} also obtained an extremely low estimate of the AME emmissivity,
a factor 20 lower than those expected from measurements on our Galaxy
\citep{planck_GP_w_ancillary_data, cbassSH2022, ameplanewidesurvey}.

In this paper, we present the data taken on M31 
with the Q-U-I JOint Tenerife Experiment Multi Frequency 
Instrument (QUIJOTE-MFI) at 11 and 13\GHz{} between 2012 and
2018, from the combination between its Wide Survey 
\citep{mfiwidesurvey} and focused raster observations.
We use these QUIJOTE-MFI data to assess the presence 
of AME in M31 following an independent procedure to that 
of \citetalias{battistelli2019} and \cite{cbassm31}. 
The paper is organized as follows. 
Section~\ref{section:data} describes the data used to
build the M31 integrated spectral energy distribution (SED)
and how they were processed.
Section~\ref{section:methodology} describes the components
assumed in the fitting of the SED of M31 and the fitting
procedure itself. Section~\ref{section:results} presents
the main results of this study, while 
Section~\ref{section:discussion} comments on the changes
for these results with different assumptions. Finally,
we present the main conclusions of this work in 
Section~\ref{section:conclusions}.

\section{Input data}
A summary of the maps used is presented in 
Table~\ref{table:maps}. All of them are used in {\tt 
HEALPix}\footnote{https://healpix.sourceforge.io}
\citep{Healpix, Healpix2} pixellization
$N_{\rm side}=512$ and smoothed to a common resolution 
of 1 degree.

\label{section:data}
\begin{table*}
    \caption{Summary of the surveys and frequency 
    maps used in this analysis. The quoted photometry
    estimates were obtained as described in 
    Section~\ref{section:AP} after applying both the
    point sources and CMB subtractions described in
    Sections~\ref{section:point_souces_subtraction}
    and \ref{section:cmb_choice} respectively. The
    photometry estimates from
    \protect\citetalias{battistelli2019}, which are
    computed using the same apertures, are shown for
    comparison.}
    \centering
    \begin{tabular}{cccccccc}
\hline
    Telescope    &  Frequency  &  Calibration  &  Resolution  &  $S_\nu$ from this work  &  $S_\nu$ from \cite{battistelli2019}  &      Reference       \\
        &  (GHz)  &  (\%)  &  (arcmin)  &  (Jy)  & (Jy)  &  \\
\hline
     \multirow{2}{*}{Various}      &       \multirow{2}{*}{0.408}       &         \multirow{2}{*}{10 }        &          \multirow{2}{*}{51}           &      \multirow{2}{*}{$13.7\pm2.9$}       &             \multirow{2}{*}{$18.4\pm1.6$}             &  \cite{haslam1982}   \\
           &              &                  &                     &             &                          &  \cite{remazeilleshaslam}   \\
    Dwingeloo    &       0.82        &         10         &          72           &      $8.6\pm1.7$       &                   -                   &  \cite{dwingeloo}   \\
      Stockert/Villa-Elisa      &       1.42        &         20         &         34.2          &      $4.9\pm1.3$       &             $5.28\pm0.41$            &  \cite{reich1982}, \cite{villaelisa}   \\
     QUIJOTE-MFI     &       11.2       &         5          &         53.2          &      $1.79\pm0.30$       &                   -                   & \cite{mfiwidesurvey} \\
     QUIJOTE-MFI     &       12.9       &         5          &         53.5          &      $1.94\pm0.36$       &                   -                   & \cite{mfiwidesurvey} \\
    WMAP    &       22.8        &         3          &         51.3          &      $1.56\pm0.14$       &             $2.00\pm0.17$            &     \cite{wmap}      \\
    \Planck{}    &       28.4        &         3          &         33.1          &      $1.36\pm0.14$       &             $1.86\pm0.15$            &    \cite{planck}     \\
    WMAP    &        33         &         3          &         39.1          &      $1.26\pm0.16$       &             $1.71\pm0.21$            &     \cite{wmap}      \\
    WMAP     &       40.7        &         3          &         30.8          &      $0.79\pm0.15$       &             $1.31\pm0.16$            &     \cite{wmap}      \\
    \Planck{}   &       44.1        &         3          &         27.9          &      $0.95\pm0.18$       &             $1.45\pm0.25$            &    \cite{planck}     \\
    WMAP    &       60.7        &         3          &          21           &      $0.84\pm0.32$       &             $1.72\pm0.42$            &     \cite{wmap}      \\
    \Planck{}    &       70.4        &         3          &         13.1          &      $1.01\pm0.27$       &             $2.12\pm0.36$            &    \cite{planck}     \\
    WMAP   &       93.5        &         3          &         14.8          &      $2.27\pm0.82$       &              $3.5\pm1.0$             &     \cite{wmap}      \\
   \Planck{}    &        143        &         3          &          7.3          &      $11.4\pm1.1$       &             $15.7\pm1.4$             &    \cite{planck}     \\
   \Planck{}    &        353        &         3          &          4.9          &     $261\pm22$     &              $318\pm24$              &    \cite{planck}     \\
   \Planck{}    &        545        &        6.1         &          4.8          &     $886\pm83$     &              $1027\pm73$             &    \cite{planck}     \\
   \Planck{}    &        857        &        6.4         &          4.6          &    $2610\pm240$    &             $3020\pm190$             &    \cite{planck}     \\
  COBE-DIRBE &       1250        &        11.6        &         37.1          &    $4680\pm610$    &       $5330\pm370$ @ 1199\,GHz       &  \cite{cobe-dirbe}   \\
  COBE-DIRBE &      2143      &        10.6        &          38           &    $6260\pm730$    &       $7020\pm230$ @ 1874\,GHz       &  \cite{cobe-dirbe}   \\
  COBE-DIRBE &       3000        &        13.5        &         38.6          &    $3010\pm460$    &       $2980\pm140$ @ 2997\,GHz       &  \cite{cobe-dirbe}   \\
\hline
\end{tabular}
    \label{table:maps}
\end{table*}

\subsection{QUIJOTE-MFI data}
\label{section:quijote-mfi}
The QUIJOTE CMB experiment~\citep{Rubino10} operates 
at the Teide Observatory (OT) of the Instituto de 
Astrofísica de Canarias (IAC), located at latitude 
$28\degr 18^{\prime}04^{\prime\prime}$ North and longitude 
$16\degr 30^{\prime}38^{\prime\prime}$ West. 
The first instrument installed on QUIJOTE was 
the Multi Frequency Instrument (MFI). This instrument
observed the sky, in intensity and polarization,
at 11, 13, 17 and 19\GHz{}, mostly in
the so-called `nominal mode' configuration, with continuous
observations at constant elevation while spinning in
azimuth. These observations constitute the Wide Survey 
(hereafter, WS), which is described in detail in 
\cite{mfiwidesurvey}. In addition to the WS data set, we
use new data obtained with raster mode observations
specifically focused on M31, intended to improve the 
sensitivity in this area.
All the analyses and results presented in this paper 
have been derived using 11 and 13\GHz{} maps.
High noise levels in the 17 and 19\GHz{} intensity maps (see 
Table~\ref{tab:sensitivity}), mostly dictated by atmospheric 
1/$f$ noise, prevent the detection of emission from M31 at these 
frequencies so  they are discarded for this analysis. 
The 11 and 13\GHz{} maps are produced combining the WS data
with the data obtained in raster scan mode around the coordinates of 
M31. A single run of the QUIJOTE Map-Making code \citep{destriper} is
required in order to optimize the recovering of the large angular
scale signal. The same strategy has been applied before \citep{
W51, hazewidesurvey} and will be used in upcoming \citep{FANwidesurvey,
perseus_raul} publications using QUIJOTE-MFI data. 

\begin{figure*}
    \centering
    \includegraphics[width=1\linewidth]{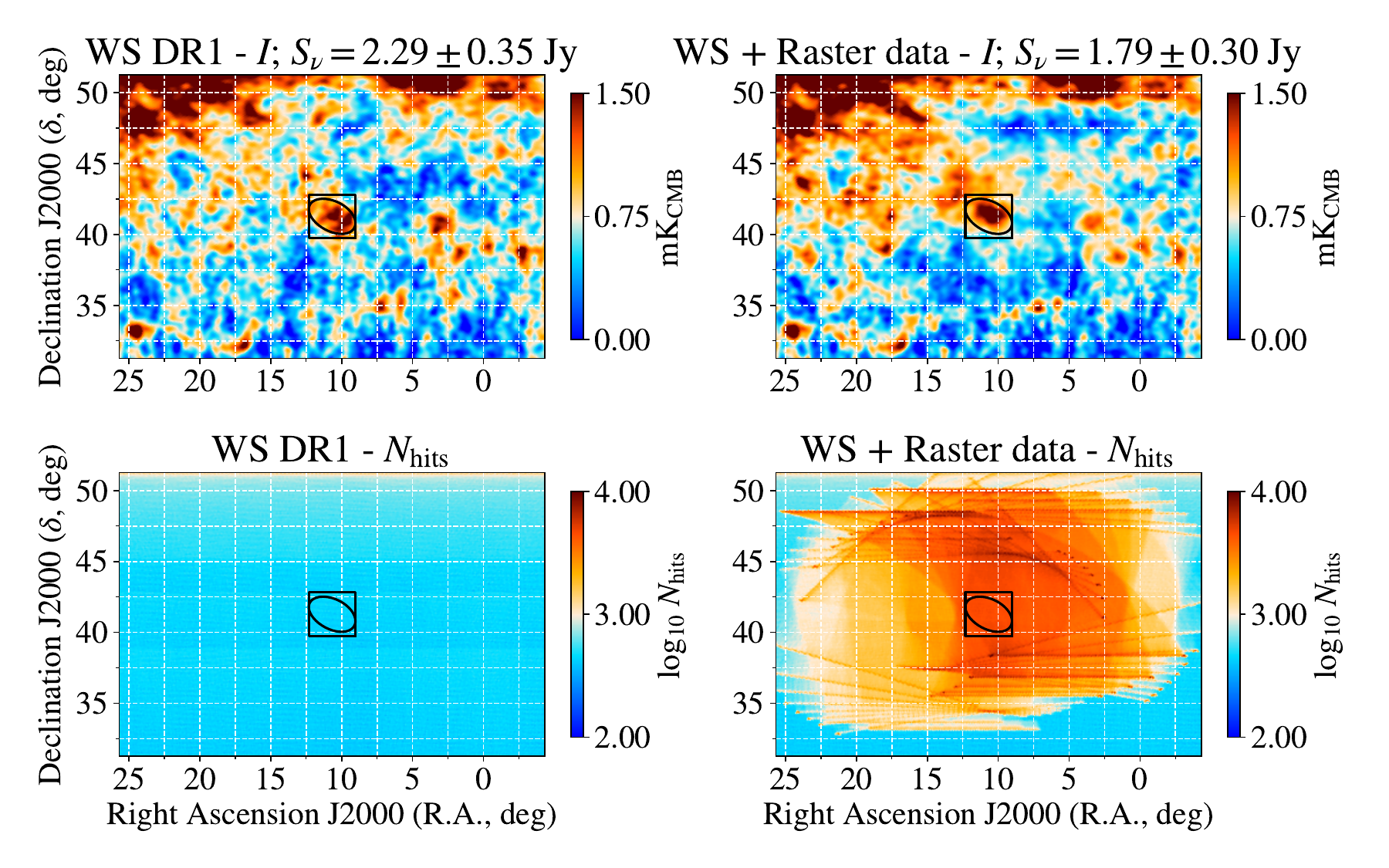}
    \caption{QUIJOTE-MFI 11\GHz{} intensity (top row)
    and $N_{\rm hits}$ (bottom row) maps for the Wide Survey (WS) 
    release (left column), and WS plus raster scan data on the M31
    field (right column). The hits are defined as 40\,ms samples in a 
    \HEALPix{} map of $N_{\rm side}=512$. The new raster data account 
    for an exposure time approximately 10 times longer than that from 
    the WS. The elliptical and 
    rectangular regions in black are those used to
    compute M31 density fluxes (and their uncertainties)
    shown in Table~\ref{table:maps} and used
    in Section~\ref{section:results}. The photometric
    results in these maps using those regions are also
    quoted in the top row maps. In the bottom row maps
    we can see how the exposure time increases by an order
    of magnitude when including the raster data with that
    from the WS.}
    \label{fig:weights_maps_comparison_WS_WS+RS}
\end{figure*}

The data in raster-scan mode were taken between May and December 2016, for a total of 931 rasters accounting for 539.1\,hours of observations (see Table~\ref{tab:raster_data}).
These raster-scans are constant-elevation azimuth scans of width 
$\approx 12^\circ/{\rm cos(EL)}$ and duration $\approx 35$~min, 
performed on different local coordinates while tracking the field, 
with elevations ranging  between $32^\circ$ and $75^\circ$. After 
projection onto the sky plane, each of these observations results in a map 
of $\approx 12^\circ$ by $\approx 12^\circ$. The combination of all these observations results in a 
sky coverage as shown in the bottom-right panel of 
Fig.~\ref{fig:weights_maps_comparison_WS_WS+RS}, where the 
footprint of the raster scans is evident. This is one of
the fields with highest integration time with the MFI. As a result,
these are the maps with the best sensitivity of all of those 
obtained with the MFI to date, both in intensity and polarization. 
Final map sensitivities, computed through a jack-knife analysis 
on the Half Mission Difference Maps (HMDM; see \citealt{
mfiwidesurvey}) are shown in Table~\ref{tab:sensitivity}: we can 
clearly see the improvement in  sensitivity
owing to the addition of raster data for most of the horns. In intensity (which are the data most of the results of this paper are based on) we reach sensitivities of 37\ and 25\,$\mu$K\,deg$^{-1}$ respectively at 11 and 13\,GHz. These are better than those from the WS, 97 and 69\,$\mu$K\,deg$^{-1}$ respectively. The improvement factor is then $\approx2.7$, slightly smaller than expected from the 10 times longer integration time. For the 
higher frequency horns, we can see that horn 2 behaves better in 
intensity while horn 4 does so in polarization, as was the case
for the WS \citep[Table 9,][]{mfiwidesurvey}.

\begin{table*}
    \caption{Sensitivity estimates computed from a 3-degree
    radius circular region around M31 on Half Mission 
    Difference Maps (HMDM), as is done in 
    \protect\cite{mfiwidesurvey}. Estimates are in units of 
$\mu$K deg$^{-1}$. The estimates from the WS
    maps alone are also shown for comparison.}
    \centering
    \begin{tabular}{cccccccc}
\hline
                    &             &  H2 17\GHz{}  &  H2 19\GHz{}  &  H3 11\GHz{}  &  H3 13\GHz{}  &  H4 17\GHz{}  &  H4 19\GHz{}  \\
\hline
 \multirow{2}{*}{I} & WS + Raster &     52.3     &     65.5     &     37.3     &     24.8     &    171.0     &    172.0     \\
                    &     WS      &    173.4     &    204.8     &     96.9     &     69.0     &    217.8     &    237.2     \\
 \multirow{2}{*}{Q} & WS + Raster &     41.2     &     54.1     &     13.4     &     11.1     &     12.4     &     10.8     \\
                    &     WS      &     62.2     &     78.2     &     45.6     &     39.1     &     36.4     &     38.9     \\
 \multirow{2}{*}{U} & WS + Raster &     42.0     &     52.9     &     13.9     &     11.3     &     11.8     &     11.1     \\
                    &     WS      &     62.0     &     77.9     &     45.6     &     39.2     &     36.4     &     39.1     \\
\hline
\end{tabular}
    \label{tab:sensitivity}
\end{table*}

We follow the same procedure and data processing pipeline that 
is currently being used in all QUIJOTE-MFI analyses dealing with
raster observations \citep{W51, hazewidesurvey}, so data are carefully 
inspected to identify and remove periods  affected by systematic effects,
in particular Radio Frequency Interference (RFI) and atmospheric 
contamination. With this aim, we produced Stokes IQU maps per horn
and frequency of each individual observation and analysed their quality, 
first by eye and then by calculating their 
noise RMS. This allows us to identify particularly bad observations,
which would degrade the final map quality. These observations are removed and not
used to produce the stacked map. However, in the case of M31, this data 
cleaning on the maps was not sufficient, since those at 11\GHz{} in particular
still presented clear signs of RFI, producing obvious striping along 
constant-declination directions. We then 
implemented an additional step consisting in combining the data 
in groups of azimuth and elevation. We then produced stacks of data for each 
group per horn and frequency by averaging the data in bins in azimuth.
In these stacks local signals clearly show up as bumps at 
specific AZ ranges. A template could then be built and removed from each 
scan, but we preferred to follow a more conservative approach and 
instead removed specific azimuth ranges of horns and frequencies that 
presented signs of RFI. On top of this, 
all observations performed at AZ$\lt 180^\circ$ and EL$\approx 32^\circ$ 
were removed as they presented particularly strong RFI that was very 
difficult to correct. All these processes resulted in removing 37\% 
of the 11\GHz{} data and 33\% of the 13~GHz data, as indicated in 
Table~\ref{tab:raster_data}. Even after this, raster data account for more
than 90\% of the data considered in the final maps.

\begin{table*}
    \caption{Description of the QUIJOTE-MFI raster scans
    dedicated to M31. Third and fourth columns refer to the 
    total amount of scans and time devoted to observe the M31
    field, respectively. The fifth column shows the fraction of
    those scans (or time) that are merged together with the data from
    the WS. The last column shows the fraction of time coming from
    the WS \protect\citep{mfiwidesurvey} over the total after adding 
    the raster data.}
    \centering
    \begin{tabular}{cccccc}
\hline
  Frequency  & \multirow{2}{*}{Observing dates}  &  Total number  &  Total observing  &  Selected time fraction  &  WS time fraction  \\
  (GHz) &  & of scans & time (h) & (per cent) & (per cent) \\
\hline
    11.2     & \multirow{2}{*}{May - December 2016} &  \multirow{2}{*}{931}   & \multirow{2}{*}{539.1} &           62.8           &        8.9         \\
    12.9     &                                       &                         &                           &           67.1           &        7.7         \\
\hline
\end{tabular}
    \label{tab:raster_data}
\end{table*}

\subsection{WMAP / \Planck{} / DIRBE}
We used the Wilkinson Microwave Anisotropy Probe 
\cite[WMAP,][]{wmap} 9-year and \Planck{} PR3 2018 maps 
\citep{planck}. We also used the zodiacal light
subtracted version of COBE Diffuse Infrared Background 
Experiment \cite[COBE-DIRBE,][]{cobe-dirbe} maps at 
240, 140 and 100\mum. WMAP and COBE-DIRBE data are 
available in the Legacy Archive for Microwave Background 
Data Analysis (LAMBDA\footnote{{\tt https://lambda.gsfc.nasa.gov}}), while \Planck{}
data are available through the \Planck{} Legacy Archive
(PLA\footnote{{\tt http://pla.esac.esa.int/pla/}}). 
The calibration uncertainties for WMAP and those \Planck{} bands 
calibrated against the CMB dipole (up to 353\GHz{} included)
are increased to 3 per cent.\footnote{\Planck{}\textit{-HFI} 545 and 
857\GHz{} bands, which are calibrated using planets, have 
calibration uncertainties greater than 6 per cent.} Note that, nominally, 
WMAP and \Planck{} data have much better (sub-percent) uncertainties. 
However, we decided
to use more conservative values to account not only for the global
gain calibration uncertainties but also for issues related with 
beam inaccuracies or colour corrections. These values are the 
same as those used in recent QUIJOTE papers \cite[e.g.][]{ 
AMEwidesurvey, W51, snrwidesurvey} or in other papers 
\cite[e.g.][]{planck2015galacticcloudsAME}. 
\Planck{} maps at 100 and 217\GHz{} are not used because of 
contamination by CO residuals.

\subsection{Other ancillary surveys}
We also used three low-frequency surveys covering the 
M31 region: the \cite{haslam1982} map at 0.408\GHz{}, 
the \cite{dwingeloo} map at 0.820\GHz{} and 
the \cite{reich1982} map at 1.42\GHz{}. We increased the
calibration uncertainty from \cite{dwingeloo} from 6
to 10\% to account for the fact that its parent angular
resolution (72\arcmin) is slightly larger than the
1-degree resolution used in this analysis. Besides, the 
\cite{reich1982} map requires applying a recalibration
factor when studying structures on the main beam scale. 
This is normally assumed to be 1.55
\cite[e.g.,][]{reich1988} when studying point sources,
but decreases when focusing on resolved objects such as M31
\citep{ameplanewidesurvey}.
Therefore, we kept the 1.3 value used by 
\cite{planckM31} and increased its calibration 
uncertainty to 20\%. We used the \cite{remazeilleshaslam} 
and \cite{CADE} reprocessed versions of the \cite{haslam1982}
and \cite{reich1982, reich1988} maps respectively. The three maps
described in this section are available in the LAMBDA
repository. 

\subsection{High resolution surveys}
Several telescopes have observed M31 with arcminute
resolution in the past. A review of these
observations is presented in \cite{berkhuijsen2003}. 
However, the 
fields covered are too small to be introduced into the 
analysis. The same happens for SRT data from 
\citetalias{battistelli2019} and \cite{fatigoni2021}:
when smoothing the data to 1 degree resolutions,
artefacts due to the presence of edges in the  
image may arise, which further complicate the analysis.
Because of this, we are not using any of these higher
resolution data in our analysis and the final set of
maps used to constrain the SEDs described in 
Section~\ref{section:methodology} is the one from
Table~\ref{table:maps}. We provide only qualitative 
comparisons to these higher resolution analyses.

\section{Methodology}
\label{section:methodology}
\subsection{SED modelling}
\label{section:foregrounds}
We have considered up to four different emission
components in this study: synchrotron, free-free, 
AME and thermal dust. These are modelled as in
\cite{ameplanewidesurvey}, with the main difference
being the lack of a component
accounting for CMB anisotropies.
Instead, we subtract the CMB anisotropies using
the {\tt NILC} \citep[see Section~\ref{section:cmb_choice},][]{
smica} map. Therefore, the flux density at frequency $\nu$ is 
described as:
\begin{equation}
\begin{split}
  S_\nu^{\rm total}(\theta) = {}&S_{\nu}^{\rm syn}(A_{\rm1\GHz{}}, 
\alphasyn) + S_\nu^{\rm ff}(\EM) \\
 &+S_{\nu}^{\rm AME}(\Aame, \nuame, \Wame) \\ 
&+S_{\nu}^{\mathrm{dust}}(\tau_{353}, \betad, \Td) \\
={}&A_{\rm1\GHz{}} \left(\frac{\nu}{1\ \mathrm{GHz}}\right)^{\alphasyn}
+ \frac{2 \kb\nu^2}{c^2} \Omega \Tff \\
&+ \Aame \mathrm{exp}\left[-\frac{1}{2\Wame^2} \ln^2\left(\frac{
  \nu}{\nuame}\right) \right] \\
&+\frac{2\h\nu^3}{c^2}  \left(\frac{\nu}{353\,\mathrm{GHz}
}\right)^{\betad}\tau_{353} 
\left(\mathrm{e}^{\h\nu/\kb \Td}-1\right)^{-1}
\Omega,
\end{split}
\end{equation}
where $A_{\rm1\GHz{}}$ stands for the synchrotron flux density
at 1\GHz{} and $\alphasyn$ for the synchrotron spectral
index; $\EM$ for the emission measure, which is used in the
definition of $\Tff$;
$\Aame$, $\nuame$ and $\Wame$ for the AME peak flux
density, frequency and width; and $\taud$,
$\betad$ and $\Td$ for the opacity, the
spectral index and the dust grain temperature, respectively, 
which describe the 
spectrum of the thermal dust emission. $\Omega$ accounts
for the solid angle of the M31 aperture (which is the
same used in \citetalias{battistelli2019}): $\Omega=4.38$\,deg$^2$.

Note that, as it has now become the standard in similar analyses, for the AME we resort to the phenomelogical model proposed by \cite{stevenson2014}. This model provides anyway a good fit to most of the spinning-dust models. For the thermal dust emission we use a single-temperature modified black-body. We are aware that in the integrated SED of M31 we mix different dust components, but even in this case this model provides a reasonably good fit. In any case, a multi-component fit would not be possible given the sparse spectral coverage.

\subsection{Aperture photometry}
\label{section:AP}
\begin{figure*}
    \centering
    \includegraphics[width=1\linewidth]{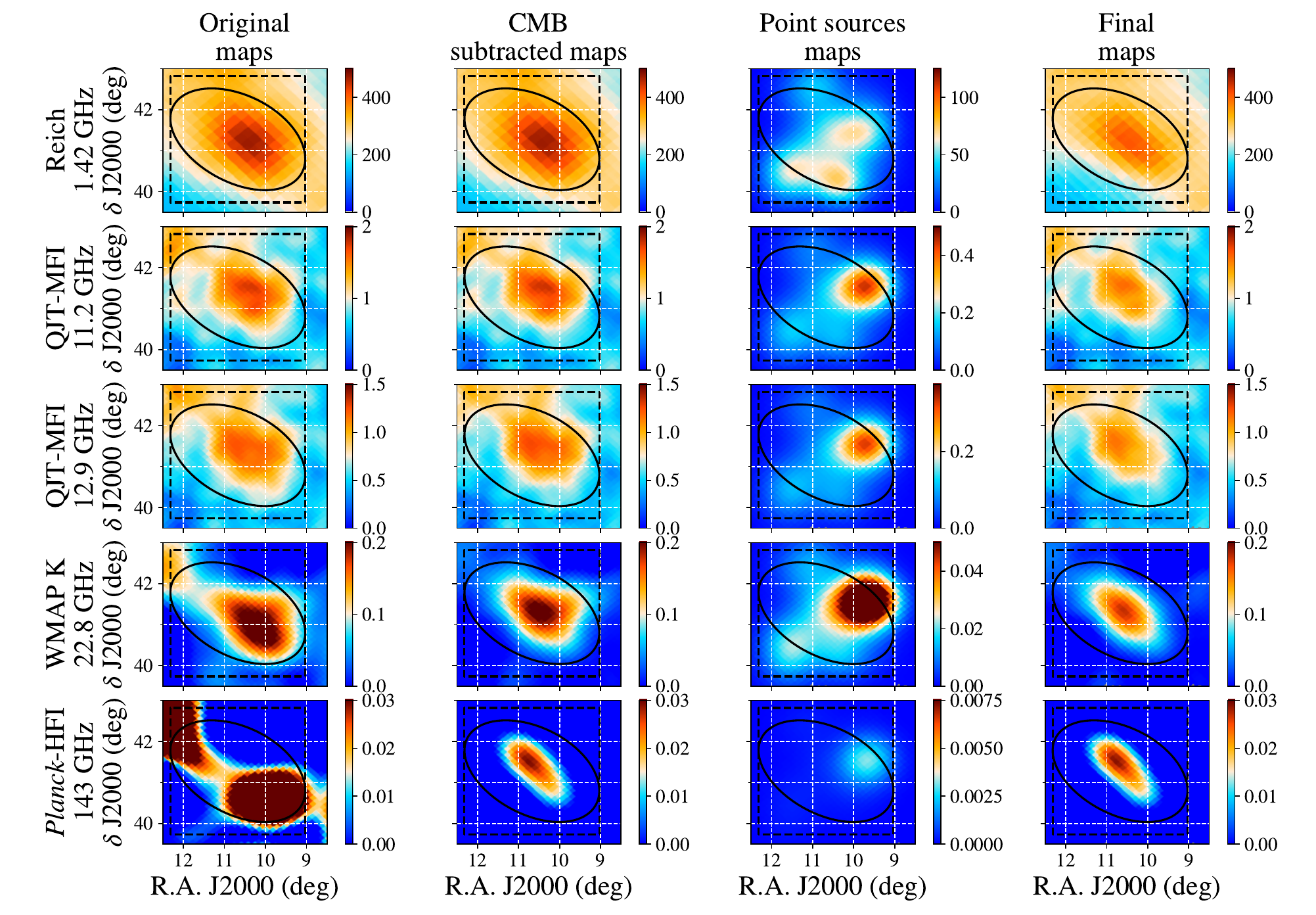}
    \caption{Examples of maps before and after CMB
    and point source subtraction. The first column 
    shows the original maps, while the second shows
    the maps after subtracting the CMB. The third
    column shows the emission from point sources
    at the frequencies of the maps. The last column 
    shows the final processed maps, which are used 
    to compute the flux densities. The different rows
    host different maps: from top to bottom, those
    are Reich at 1.42\GHz, QUIJOTE-MFI at 11.2\GHz{}
    and 12.9\GHz{},
    WMAP at 22.8\GHz{} and \textit{Planck-HFI} at 
    143\GHz. The elliptical aperture used to extract
    M31 flux is shown as a solid line. The 
    rectangular region marked by the dashed 
    line is used to compute the background emission
    level. All maps have mK units. The colourbar levels
    for the point sources maps are fixed at 25\% of the 
    amplitude of the colourbar in the original maps.
    In WMAP K and MFI 11\GHz{} maps, it is clear how
    B3 0035+413 (Section~\ref{section:3B0035})
    emission distorts the elliptical morphology of M31
    as seen in SRT (Fig.~\ref{fig:m31_by_srt}) or
    thermal dust (143\GHz{}, bottom row) maps. This
    morphology is recovered after subtracting the
    source, as can be seen in the last column. The 
    most important correction for the
    143\GHz{} map, however,  is the CMB anisotropy subtraction.}
    \label{fig:maps_before_after_subtraction}
\end{figure*}

We adopt the same M31 aperture used in \citetalias{battistelli2019}:
an ellipse with semi-axes 91.5\arcmin and 
59.5\arcmin, position angle $-$52\degr (east-to-north)
and centred on (RAJ2000, DecJ2000) = (10.68\degr, 
41.27\degr).\footnote{\tt https://simbad.u-strasbg.fr/simbad/sim-basic?Ident=M31\&submit=SIMBAD+search}
Our background aperture is defined as those pixels
also present in the \citetalias{battistelli2019} SRT image
outside the M31 ellipse: this defines a rectangle with 
edges located at ${\rm RAJ2000}=(9.15\degr,12.25\degr)$ and 
${\rm DECJ2000}=(39.74\degr, 42.82\degr)$.
We decided to keep the source
and background regions identical to those in 
\citetalias{battistelli2019} not only to provide a direct comparison
with that study, but also to be able to use the point source
catalogue released later by \cite{fatigoni2021}. Both M31 
and background apertures can be seen in 
Fig.~\ref{fig:maps_before_after_subtraction}.
We computed M31 flux densities as in previous studies
\cite[e.g.][]{jarm_poldebiased, W44, ameplanewidesurvey} by
using the mean and the median values within the source and 
background apertures respectively (Eq.~\ref{eq:AP}). 
The uncertainty for these flux densities are computed from the 
standard deviation within the background aperture 
(Eq.~\ref{eq:AP_unc}):
\begin{equation}
    S_\nu = a(\nu)\Omega T = \frac{2k_{\rm B}\nu^2}{c^2}\frac{x^2
    \mathrm{e}^x}{(\mathrm{e}^x-1)^2} \Omega \left[\overline{T}_{
    \rm aper} - \mathrm{med}(T_{\rm BG})\right]
\label{eq:AP}
\end{equation}
\begin{equation}
    \sigma_{\rm AP} = a(\nu)\Omega\,\sigma(T_{\rm BG})\sqrt{
    \frac{\nbeam}{n_{\rm aper}} + \frac{\pi}{2}\frac{\nbeam
    }{n_{\rm BG}}}
\label{eq:AP_unc}
\end{equation}
where $\nbeam$ is the number of pixels within a one 1 degree
beam, while $n_{\rm aper}$ and $n_{\rm BG}$ are the number of pixels
within the M31 and background apertures, respectively.
We add this factor because of the assumption that the noise 
is completely correlated on beam scales: this is a conservative way 
of estimating the uncertainties for aperture photometry studies
\citep{Perseus, W44}. Finally, the calibration uncertainty
quoted in Table~\ref{table:maps} is added quadratically to the 
uncertainty estimation:
\begin{equation}
\sigma_{S_\nu} = \sqrt{\sigma_{\rm AP}^2 + \mathrm{cal}^2\cdot S_\nu^2},
\label{eq:sigma_AP_unc_calibration}
\end{equation}

\subsection{MCMC}
\label{section:MCMC}
We ran several Maximum Likelihood Estimator (MLE) analyses
using Markov Chain Monte Carlo (MCMC) samplers
for various scenarios, using {\tt emcee}
\citep{emcee}. Each of these MCMCs
ran for a fixed length of $N=10^5$ steps.
Once they ended, the convergence of the chains was
assessed by computing their autocorrelation 
times, $\tau$. The chains have converged if
$\tau<N/50$: all cases fullfilled this
requirement. Colour corrections ($\textbf{cc}$) are 
embedded within the MCMC log-likelihood ($\log\mathcal{L}$) 
and computed iteratively, using
{\tt fastcc} and {\tt interpcc} from \cite{fastcc}, as
in the following equation:
\begin{equation}
\begin{split}
    \log\mathcal{L} =  
-0.5 \cdot \left(\textbf{\textit{S}} - 
\frac{\textbf{\textit{S}}^{\rm total}(\theta)}{\textbf{\textit{cc}}}\right)^{\rm T}\textbf{\textit{C}}^{-1}\left(\textbf{\textit{S}} - \frac{\textbf{\textit{S}}^{\rm total}(\theta)}{\textbf{\textit{cc}}}\right),
\label{eq:chi2_w_CC}
\end{split}
\end{equation}
where $\textbf{\textit{S}}$ and $\textbf{\textit{S}}^{
\rm total}(\theta)$ stand for the measured and expected 
flux densities.
$\textbf{\textit{C}}$ is the covariance matrix between the surveys,
which has all its off-diagonal elements equal to zero except 
for those corresponding to the 11 and 13\GHz{} maps, whose 
noise is partially correlated. Further details on this 
methodology are available in Section~3.3.2 of  
\cite{ameplanewidesurvey}. We use the same flat minimal priors
on the parameter posteriors as in that study.

\subsection{Point source subtraction}
\label{section:point_souces_subtraction}
\begin{figure*}
    \centering
    \includegraphics[trim=0cm 2.5cm 0cm 1cm, width=1\linewidth]{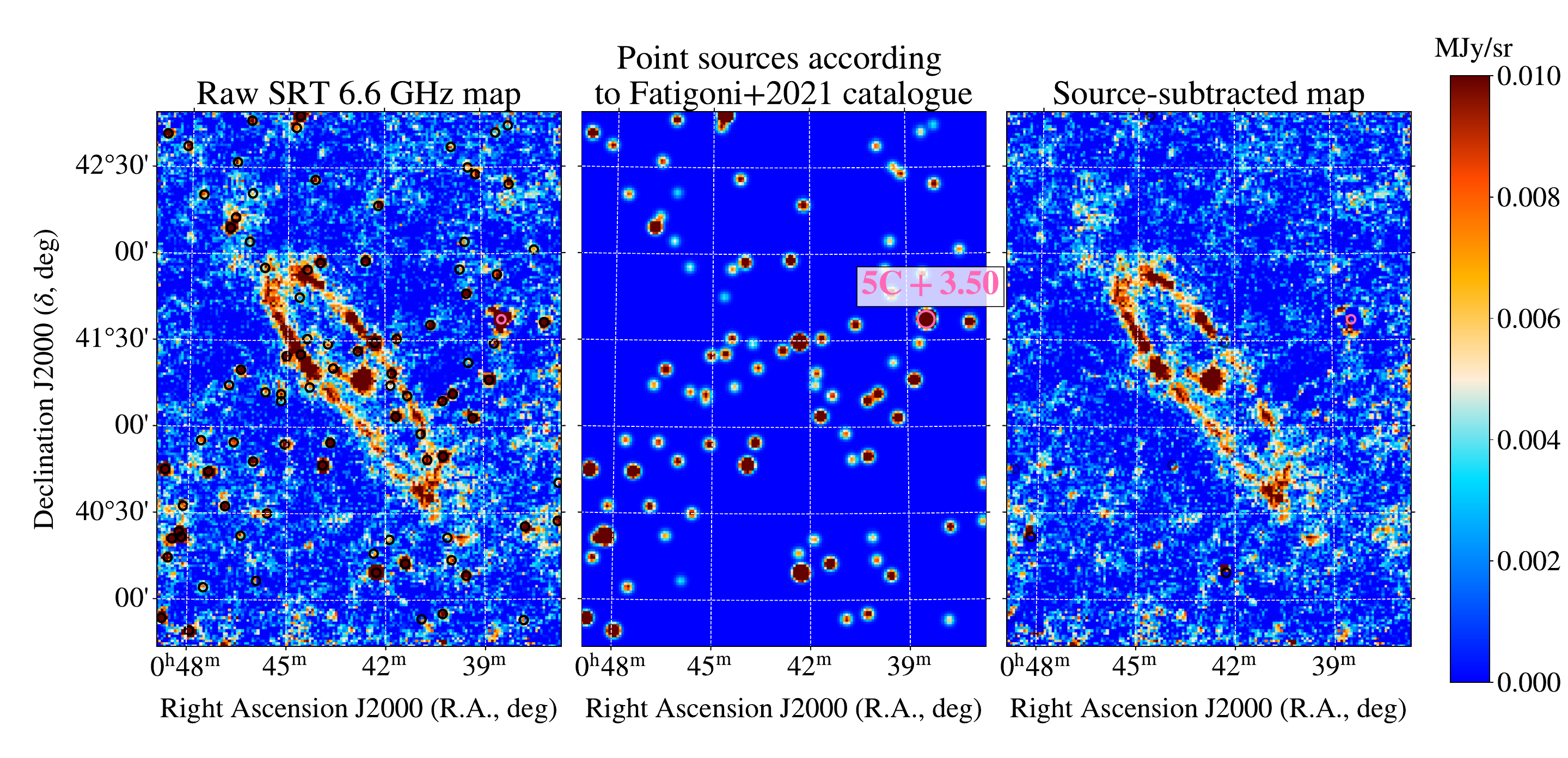}
    \caption{Left: M31 map obtained with the SRT
    at 6.6\GHz{} before subtracting point sources. 
    Point sources are indicated with black 
    circles. Middle: point sources map at 6.6\GHz{}
    SRT resolution (2.9$\arcmin$), according to 
    \protect\cite{fatigoni2021}. Right: 
    M31 map after subtracting the point sources. 
    In this last map, the opacity for the point 
    source markers is directly proportional to 
    their amplitude at 6.6\GHz{}.}
    \label{fig:m31_by_srt}
\end{figure*}

We used the catalogue from \cite{fatigoni2021}\footnote{
{\tt https://cdsarc.cds.unistra.fr/viz-bin/cat/J/A+A/651/A98}}
to subtract the point sources within the SRT M31 
observed field (Fig.~\ref{fig:m31_by_srt}). The
catalogue provides the flux density at $\nu_{\rm ref}
=1\GHz{}$ ($A$) plus a constant spectral
index ($\alpha$) and a curvature parameter ($k$), so 
the flux density at a frequency $\nu$ is defined as:
\begin{equation}
    S = A \left(\frac{\nu}{\nu_{\rm ref}}\right)^{\alpha}
    \exp\left[k\left(\frac{\nu}{\nu_{\rm ref}}\right)^{-1}\right].
\label{eq:flux_modelling}
\end{equation}

\subsubsection{B3 0035+413 variability}
\label{section:3B0035}
One of the point sources, B3 0035+413 (or 5C+3.50), 
required a detailed analysis for two reasons.
First, its spectral index is one of the flattest in the 
full catalogue, with $\alpha=-0.08$,\footnote{No uncertainties
are provided in the catalogue. \cite{cbassm31} also found a 
significantly flat spectrum ($\alpha=-0.04\pm0.01$) at 
radio-to-microwave frequencies. Variable extragalactic sources 
tend to have flatter spectra \citep{flatterspectravariability1, 
flatterspectravariability2} because of optically thick synchrotron
emission. This requires high surface brightness, which for the 
range of flux densities that we actually measure, implies small
angular sizes. On the other hand, variability on yearly timescales 
require small light-crossing timescales, or sizes. A source with
a steep spectrum has more flux coming from their outer regions,
thus their light-crossing timescale would be large and 
variability could not be measured.} compared 
to the median value $\alpha=-0.78$ across all sources.
This, together with its high flux density, 
implies that this source remains bright for many
of the frequency bands studied. It has a null curvature ($k=0$) 
estimate. It is clear in 
Fig.~\ref{fig:maps_before_after_subtraction}
how the morphology of M31 in the WMAP K map 
resembles the ellipse seen in SRT or thermal dust data
only once the emission from this source is subtracted.
Second, this is a variable source:
excess emission was recorded by the 40 m telescope
at the Owens Valley Radio Observatory (OVRO) at 15\GHz{}
between 2009 and 2012 \citep{OVRO}, increasing from 0.45 to
0.55\Jy{}. This excess overlaps in time with \Planck{}
observations and therefore accounts for 20\% of its
flux before the event.
It is impossible to know the exact behaviour of
this source over the whole time span covered by 
the surveys depicted in Table~\ref{table:maps}
(from the '80s well into the 2010s). Besides,
variations can be limited to only part
of the source spectra. Therefore, we decided to estimate 
the change in the photometric results induced by large 
variations (worst case scenario) of this source flux, 
and increase the uncertainties accordingly.

We ran a set of $10^4$ simulations for this source, 
defining its amplitude at 6.6\GHz{} from a random 
Gaussian realization assuming a $\sigma$ equal to 50\%
of the measured flux in the \cite{fatigoni2021} catalogue.
We assumed this variability level after the analysis ran in Section 4.2 of \cite{cbassm31}, where the variability of this source was found to be between 30 and 50\% from yearly \Planck{} 28.4\GHz{} maps. We used the 5C+3.50 broken power spectral model from \cite{cbassm31} to build an estimate for this source intensity
at each map frequency. This estimate is then projected onto a map
for each frequency assuming the SRT FWHM beam. These maps
are smoothed to 1 degree resolution and transformed
to \HEALPix{} maps. These maps, which contain only the signal
from 5C+3.50, are then subtracted to the frequency
maps. We performed aperture photometry on these final maps,
as explained in Section~\ref{section:AP}, and assessed
the changes in the extracted flux density for the full M31
aperture. We found that the introduced 50\% $\sigma$
variability level translates into a maximum 40\% variation 
in the computed flux density of M31 at \Planck{} 28.4\GHz{} band. 
We increased the flux density error at each frequency by adding in quadrature the uncertainty associated with the variability of this source. Therefore, Eq.~\ref{eq:sigma_AP_unc_calibration} transforms into:
\begin{equation}
\sigma_{S_\nu} = \sqrt{\sigma_{\rm AP}^2 + \mathrm{cal}^2\cdot S_\nu^2
 + (\sigma^{\rm var}_\nu)^2\cdot S_\nu^2},
\label{eq:sigma_AP_calibration_variability}
\end{equation}
where $\sigma^{\rm var}_\nu$ is the variation level found for each band.
It is only larger than $10\%$ for WMAP and \Planck{} bands between 20
and 50\GHz{}, while being negligible for low ($\nu<10\GHz{}$) and high 
($\nu>90\GHz{}$) frequency bands.


\subsection{Subtraction of CMB anisotropies}
\label{section:cmb_choice}
The contribution of CMB anisotropies to the global SED of M31 is 
not negligible, as can be seen in the \Planck{} 143\GHz{}
maps in the bottom row of Fig.~\ref{fig:maps_before_after_subtraction}. 
In fact, given the relatively large angular size of M31 and its low 
flux density, CMB anisotropies make an important contribution to the
uncertainty in the measured flux densities. As mentioned in 
Section~\ref{section:data}, we used 1-degree smoothed maps in 
this study: at that scale, CMB anisotropies show an RMS of 
$\approx80\,\mu$K. When translated into flux density, that uncertainty value 
dominates over the dispersion of the rest of the emission from the 
background region. It is thus preferable first to subtract the CMB
anisotropies and then fit for the other components at the SED
level. Several maps tracing CMB anisotropies exist, but
we focused on the four official \Planck{} collaboration CMB maps,
from SMICA, NILC, SEVEM and COMMANDER methods \citep{smica}.

In order to study to what degree the results presented in this paper
could be affected by the selection of the CMB component-separated map,
we performed aperture photometry on all CMB maps.
First, we used SMICA, following the methodology of \citetalias{battistelli2019}. 
We found that the SMICA CMB integrated contribution in the M31 region
was the most important of the four maps ($+1.0\,\mu$K). 
This estimate (and those
for the the other maps quoted later on) was computed as the mean 
temperature within the M31 aperture minus the median of the background 
region, using the apertures from 
Section~\ref{section:AP}. COMMANDER is the opposite case: its estimate 
is the least important of the four ($-2.4\,\mu$K). 
NILC and SEVEM yielded similar CMB mean temperatures 
($-1.3\,\mu$K and $-0.8\,\mu$K), which sit between 
the SMICA and COMMANDER values as good trade-off values: we
decided to use NILC after performing a visual inspection of the maps 
(there are visible residuals in SEVEM on the M31 outer ring). 
Using any of the extreme values
(SMICA or COMMANDER) biases the AME estimation to high or low values.
This is because the amplitude from the CMB maps within the M31 aperture 
is comparable to the one from the
emission coming from the galaxy itself. 
This is further discussed in 
Section~\ref{section:ame_dependence_on_cmb_map}.
Finally, in order to reflect uncertainties 
associated with the subtraction of the CMB anisotropies, the error bar 
derived from Eq.~\ref{eq:AP_unc} was increased by adding in quadrature the
standard deviation of the integrated CMB anisotropy
estimates from the previous maps (1.23\,$\mu$K). In this way,
the error budget of the flux density estimates increases mostly in the 
spectral range where the CMB contribution is important.

\section{Results}
\label{section:results}
\begin{figure*}
    \centering
    \includegraphics[trim= 0cm 0cm 0cm 0cm, width=0.75\linewidth]{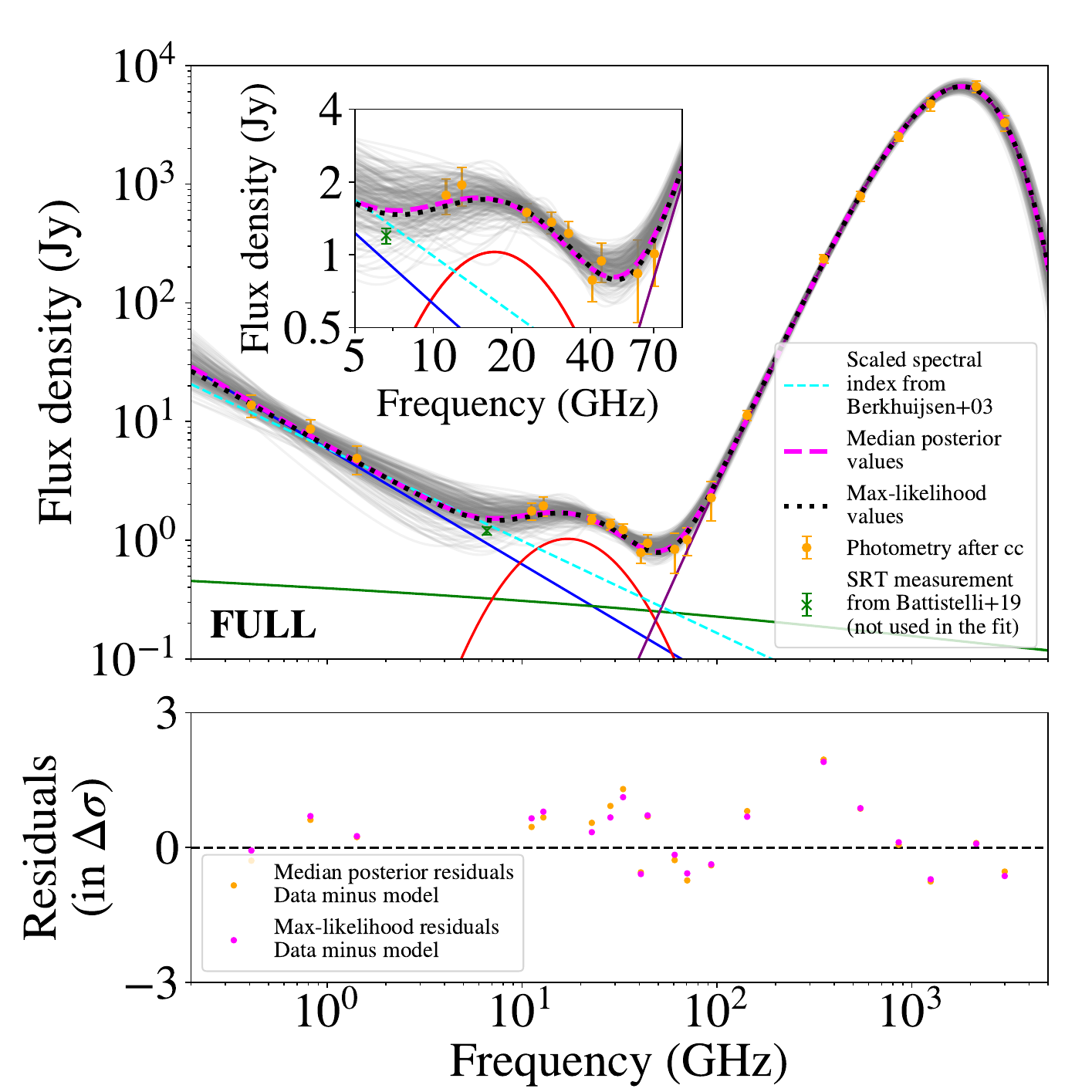}
    \caption{Top: fitted SED in M31, considering all
    components (FULL scenario) defined in 
    Section~\ref{section:foregrounds}. Each of the solid
    lines account for an emission component (blue: synchrotron, 
    green: free--free, red: AME, purple: thermal dust).
    Grey lines show different realizations of the 
    MCMC. Both fits 
    obtained using the median values from the parameters
    1D marginal distributions and the parameter combination
    returning the maximum likelihood estimate are plotted.
    Photometry estimates are plotted before and after
    applying colour corrections (cc); values for the
    no-cc case are given in Table~\ref{table:maps}.
    The parameters describing the different cases are
    shown in Table~\ref{table:SED_cases}. The scaled 
    emission from \protect\cite{berkhuijsen2003} shown
    here is calculated using the spectral index of the
    full emission,
    not the one for synchrotron alone. The photometry
    estimate from \citetalias{battistelli2019} at 
    6.6\GHz{} is shown for comparison only:
    it was not used in this fit or in any of the 
    following, although its estimate is consistent
    with our model estimate at that frequency. Bottom: 
    photometry residuals for the fits using either the
    median values from the 1D posteriors or the maximum
    likelihood combination for the parameters.}
    \label{fig:SED_cases_1}
\end{figure*}

We recover the AME peak flux density parameter, $\Aame$,
in M31 with a $3.2\sigma$ significance,
as can be seen in Table~\ref{table:SED_cases} and 
Fig.~\ref{fig:SED_cases_1}: $\Aame=1.03\pm0.32\Jy{}$.
At 25\GHz{}, we estimate the AME flux density to be
$0.83\pm0.32\Jy{}$, 1.7$\sigma$ below the 
value of  $1.45\pm0.18$\Jy{} quoted by \citetalias{battistelli2019}. 
The reason for this difference could be that our 
measured flux densities between 10 and 60\GHz{} (where AME 
is expected to be most important) are lower than those
from \citetalias{battistelli2019}, probably due to differences
in the filtering of the ancillary data together with the
different choices for the CMB map (\citetalias{battistelli2019}
used SMICA while we are using NILC, as further discussed in 
Sections~\ref{section:cmb_choice} and 
\ref{section:ame_dependence_on_cmb_map}). For example, our photometry
estimate for \Planck{} 28.4\GHz{} band is 74\% than that from 
\citetalias{battistelli2019}: correcting for 
that factor, our $\Same$ estimates at 25\GHz{}
would increase to $1.1\Jy{}$, closer to those from
\citetalias{battistelli2019}. The corner plot showing
the parameter posteriors used to build the previous SED
is shown in Fig.~\ref{fig:corner_plot_full}.

Our fitted synchrotron spectral index, $\alphasyn=-0.97_{-0.23}^{+0.18}$ is consistent with those obtained by \cite{berkhuijsen2003},
 $-1.0\pm0.2$, and \cite{planckM31}, $-0.92\pm0.16$.
It is also consistent with the value obtained by \citetalias{battistelli2019},
$\alphasyn=-1.1\pm0.1$, owing to large uncertainty in this
parameter due to unresolved parameter degeneracies. This is because of the lack 
of data with good calibration uncertainties below 10\GHz{}.
It is worth noting that, except for the thermal dust spectral index, the differences between our best-fit parameters and those of \citetalias{battistelli2019} are always within the $1\sigma$ level. The study of \citetalias{battistelli2019} benefitted from the addition 
of the SRT data point at 6.6\GHz{}, which is critical to alleviate the 
degeneracy between the synchrotron and free--free components. The value
of our fit at 6.6\GHz{} is not too different from the one quoted in the
analysis of \citetalias{battistelli2019}: our flux density estimate
value is $1.68^{+0.36}_{-0.34}$\Jy{} compared to $1.199\pm0.087$\Jy{} from 
\citetalias{battistelli2019}. The two are only $1.3\sigma$ away.

Degeneracy with synchrotron emission prevents a precise determination of the level of free--free emission from our compoment-separation analysis. Actually, Fig.~\ref{fig:corner_plot_full} clearly shows that the 1D posterior of EM is consistent with a non-detection of free-free emission, and an upper limit would probably be more appropriate. The 68\% upper limit on the free-free flux density at 1~GHz derived from this posterior is 0.52\Jy{}. This is consistent with the expected value derived from the H$\alpha$ emission. By integration on the map of \cite{finkbeiner2003} with the aperture and background annulus described in Section~\ref{section:AP}, and using the formalism explained in \cite{dickinson2003}, we derive an expected free--free flux density at 1\,GHz of $S_{\rm 1~GHz}^{\rm ff}=0.15\pm 0.02$\,Jy. Due to absorption of H$\alpha$ emission this value might be regarded as a lower limit. Absorption correction suffers from important uncertainties, related for instance with the way absorption measured at other wavelengths is translated into H$\alpha$ absorption. Using the formalism described in \cite{dickinson2003} and the reddening $E(B-V)$ map derived by the Planck collaboration \citep{cpp2013-11} we get $S_{\rm 1~GHz}^{\rm ff}=0.19$~Jy. If we instead use the reddening $A_{\rm V}$ map, also derived by the Planck collaboration \citep{pir29} we get $S_{\rm 1~GHz}^{\rm ff}=0.27$~Jy. These values are both consistent with our derived upper limit.

Another common way of estimating free-free emission relies on measurements of the star formation rate (SFR). \cite{ford2013} measured $0.25\pm 0.05\,M_\odot$/yr from UV and 24-micron data. 
Equation~6 of \cite{murphy2012} provides a relation between SFR and spectral luminosity. Using this equation and assuming an average electron temperature $T_{\rm e}=8000$~K (typical value from our Galaxy) we get $L_\nu=4.92\times 10^{19}$\,W/Hz at 1\,GHz. From this, and using a distance to M31 of $d=785$\,pc \citep{mcconnachie2005} we get a flux density at 1~GHz of $S_{\rm 1~GHz}^{\rm ff}=0.67$\,Jy, a value that is roughly consistent with our derived upper limit (we have intentionally omitted the statistical error in this estimate as systematic errors associated with the scaling relations used are expected to be larger).
 
Thus, given that the estimate for the amplitude of free--free is consistent with zero, we repeated our analysis
without taking this component into account. The absence of a
free--free contribution is compensated 
by a slightly wider AME component: $\Wame=0.66\pm0.18$
now, as compared to $0.58\pm0.16$ when free--free 
was considered.\footnote{In fact, we can see in the corner
plot (Fig.~\ref{fig:corner_plot_full}) that the free--free
amplitude and $\Wame$ are slightly anticorrelated. Large values
of $\Wame$ imply wide AME distributions that could 
resemble power-law distributions similar to free--free
between 10 and 60\GHz{}, where AME is most important.}
However, most AME theory models 
\citep[\texttt{SPDUST},][]{spdust1,spdust2} expect values
for $\Wame$ smaller than 0.6. Previous studies have 
resolved this issue by adding a second AME component (e.g.
\citealp{planck2016Xforegroundmaps}). However, in
the case of the integrated emission
from a galaxy like ours, we expect a large number of AME 
emitting regions with different spectral parameters
within our aperture, leading to a broadening of the
integrated AME distribution. 
We tried to fit a second AME component, but found that
the MCMC chains stopped converging and the parameters defining 
this new component were not properly constrained. 
In conclusion, we decided to use as the default model in
this analysis the one with free--free emission, even 
though its $\chisqred$ value is slightly higher than
the value when no free--free is considered ($0.39$ vs.\ $0.61$,
$\chisqred$ being systematically smaller than 1 most probably
owing to an overestimation of the photometry uncertainties).
This decision is based on the fact that we know, from
different kinds of observational evidence (e.g. the presence of star
formation or H$\alpha$ emission), that free--free emission 
originating in M31 must exist.

\subsection{Statistical evaluation of the presence of AME
in M31}
\begin{figure*}
    \centering
    \includegraphics[width=1\linewidth]{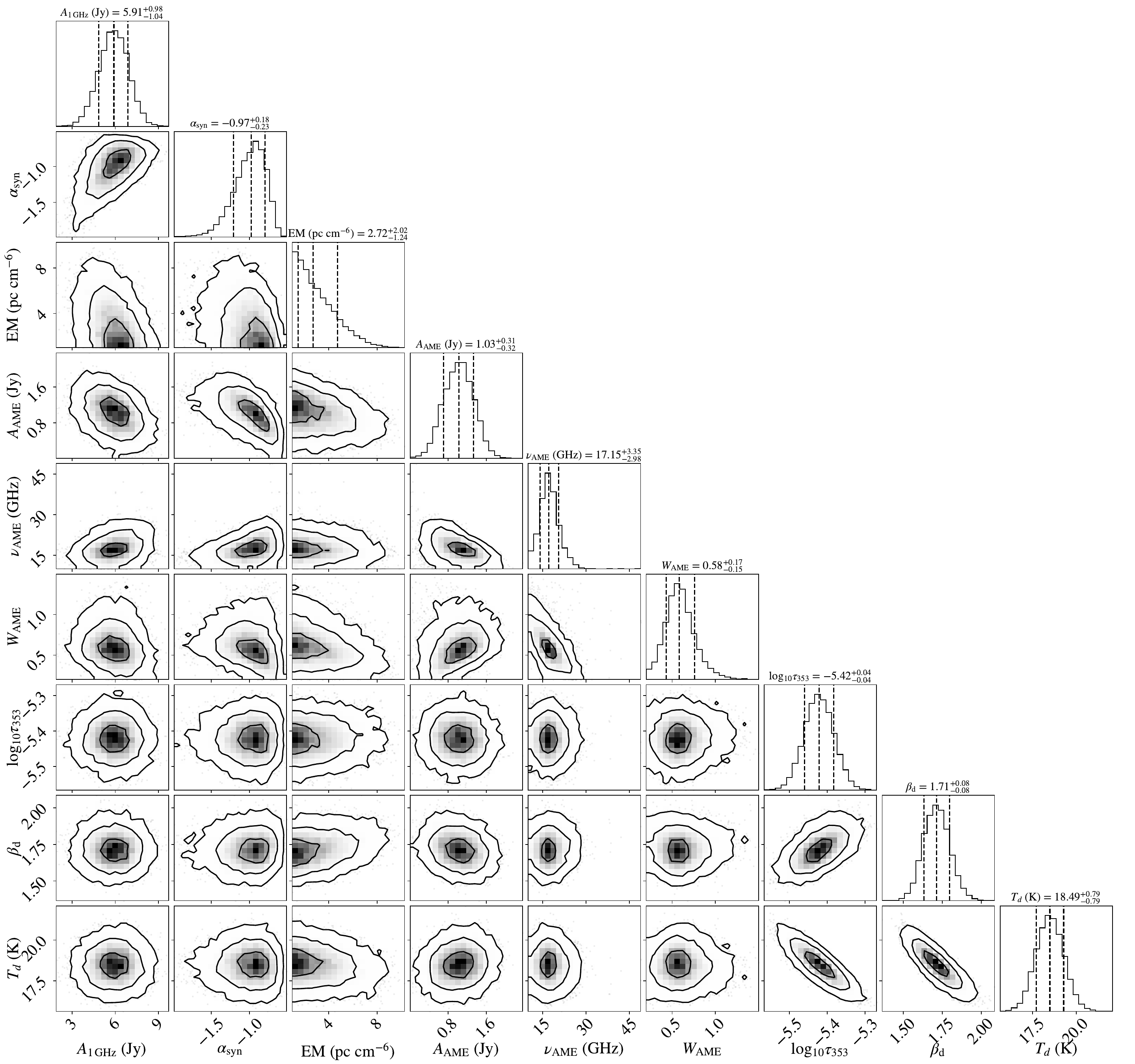}
    \caption{Corner plot example, from the case
    considering all emission components (i.e.\ 
    the one also shown in Fig.~\ref{fig:SED_cases_1}).
    We can see how the most important degeneracies
    (aside from the one between the thermal dust components,
    which are related to the modelling) are those between
    the AME amplitude ($\Aame$) and the free--free one
    ($\EM$), but $\EM$ is also degenerate with
    $\Wame$. This degeneracy arises because a wide enough
    AME component would 
    resemble the behaviour of a flat power law
    (similar to a free--free component) for those
    frequencies where free--free is most important
    (between 5 and 100\GHz{}). Solid line contours encompass
    1, 2, and 3$\sigma$ levels.}
    \label{fig:corner_plot_full}
\end{figure*}

We repeated the analysis without
the AME component to ensure that AME in M31 could not be
mistaken for free--free emission. The $\chisqred$
value obtained in that case is at least a factor 2.5
worse than any of the two other scenarios where an
AME component is introduced (Table~\ref{table:SED_cases}). 
$\chi^2_{\rm red}$ values are computed 
using the median values from the distributions of the
parameter posteriors shown in Fig.~\ref{fig:corner_plot_full}.
These median values are those quoted in the table. The worst
fit is the one for which no AME component is considered.

We used the Akaike information 
criterion \citep[AIC,][]{AIC} and the Bayesian information 
criterion \citep[BIC,][]{BIC} to compare the goodness of the
fits with and without AME. This was done in order to account
for the non-Gaussianity of the likelihood, mostly due to the
presence of non-linear parameters in the SED fitting (e.g.\
the synchrotron spectral index). AIC and BIC are computed
as follows:
\begin{equation}
\begin{split}
    \AIC &= 2k - 2\log\mathcal{L} \\
    \BIC &= k\log(n) - 2\log\mathcal{L},
\end{split}
\end{equation}
where $k$ stands for the number of fitted parameters, while $n$
is the number of data points introduced in the fit. $\log\mathcal{L}$
is the maximum log-likelihood value computed within the MCMC,
as explained in Section~\ref{section:MCMC} and 
Eq.~\ref{eq:chi2_w_CC}. Larger differences on the AIC or BIC estimates 
from two models will imply a heavier preference of the data for the 
model with the lowest value. The main difference
between the two is that the BIC penalizes the addition of 
parameters more heavily than the AIC. We compare the case with
all the components and the case without AME, as we 
previously explained that the case with no free--free component
is not physical:
\[
\begin{split}
  &\AIC_{\rm all} = 23.6; \quad \AIC_{\rm no\,AME} = 31.9 \\
  &\BIC_{\rm all} = 32.5; \quad \BIC_{\rm no\,AME} = 37.9.
\end{split}
\]
According to the AIC, the relative likelihood of the model 
without an AME component compared to the model with all
components considered is:
\[
\exp((\AIC_{\rm all} - \AIC_{\rm no\,AME})/2) = 0.0016
\]
and the model with AME is strongly preferred. When
using the BIC, we found the relative likelihood of the model
without AME to be:
\[
\exp((\BIC_{\rm all} - \BIC_{\rm no\,AME})/2) = 0.0067,
\]
over a factor 4 larger. This is consistent with BIC penalizing
the larger number of parameters considered in the case with
all components, compared to that without AME. However, the
model without AME is still below 0.01 times as probable as that
with AME, pointing to a strong preference for the latter.

\begin{figure}
    \centering
    \includegraphics[width=1\linewidth]{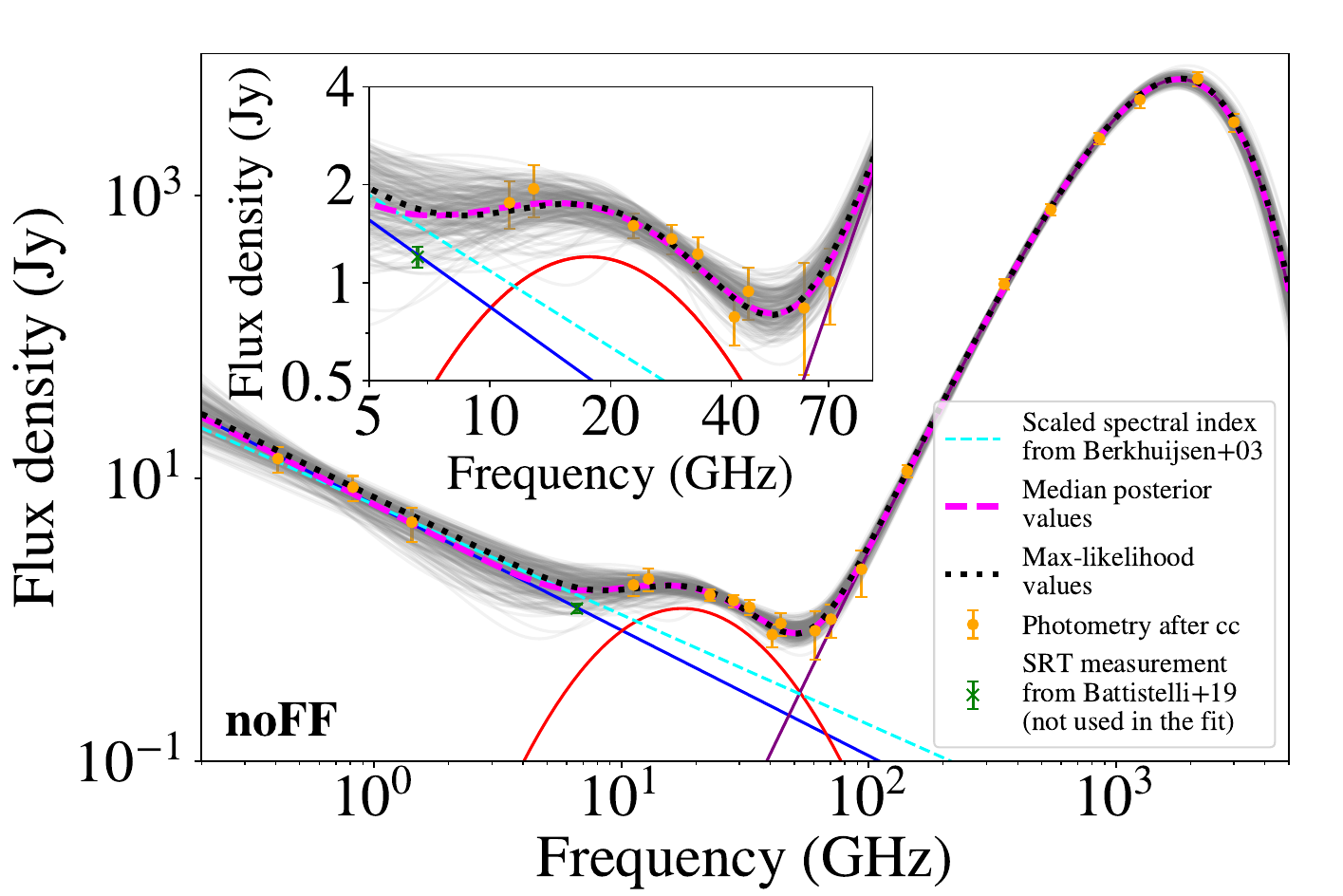}
    \includegraphics[width=1\linewidth]{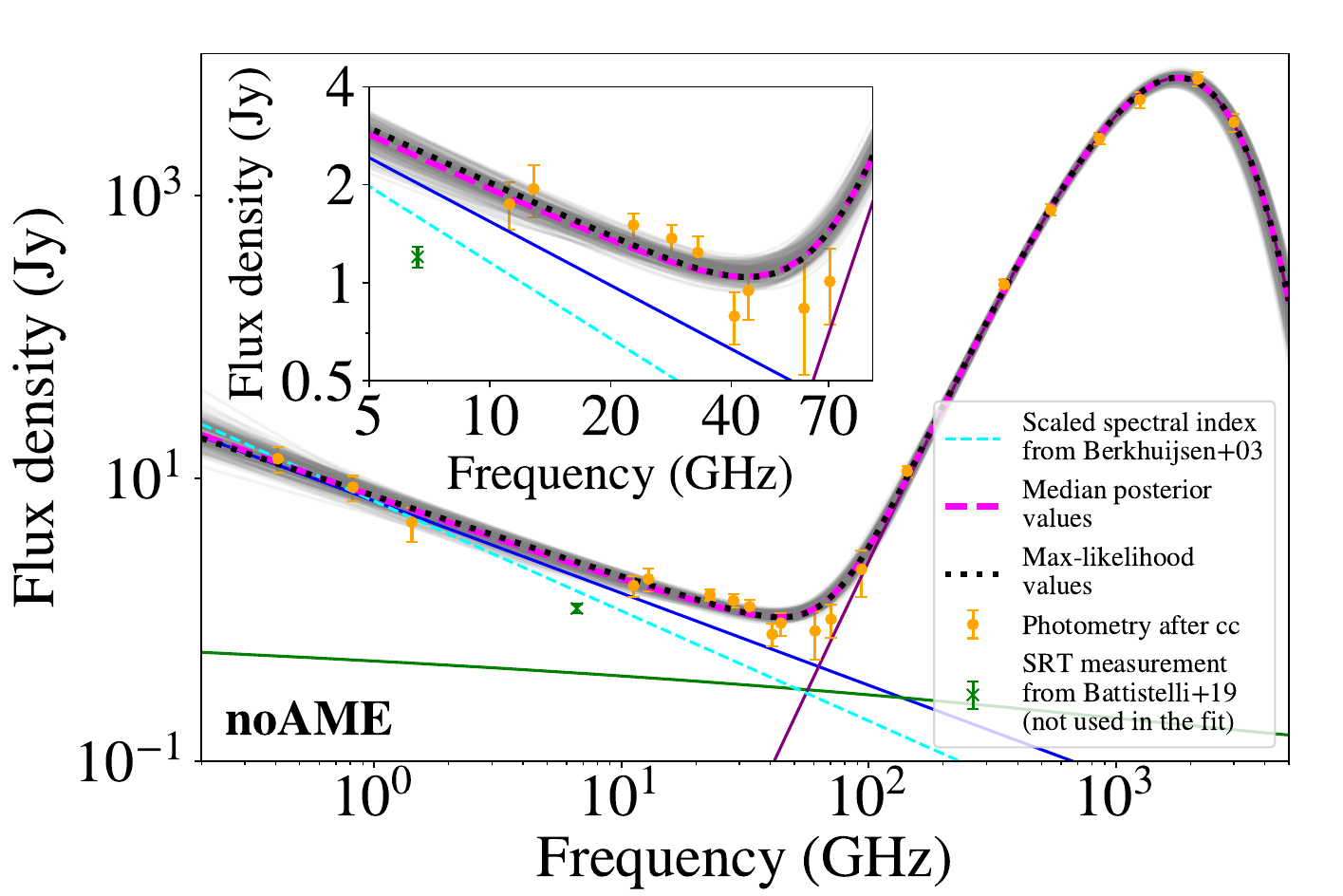}
    \caption{Same M31 SED as in Fig.~\ref{fig:SED_cases_1},
    but now without a free--free component (top, noFF), and 
    finally without an AME component (bottom panel, noAME).
    The parameters describing the different cases are shown
    in Table~\ref{table:SED_cases}.}
    \label{fig:SED_cases_2}
\end{figure}

\begin{table*}
    \caption{Comparison between the results for the
    different cases considered, shown in Figures
    \ref{fig:SED_cases_1} and \ref{fig:SED_cases_2}.
    These values are obtained as the median values 
    from the marginalised 1D distributions, as shown in
    Fig.~\ref{fig:corner_plot_full}.}
    \centering 
    \begin{tabular}{ccccc}
\hline
                                     &  Case 1 (all components)  &  Case 2 (no AME)  &  Case 3 (no free-free)  &  From \cite{battistelli2019}  \\
\hline
        $S_{\rm 1\,GHz}$ (Jy)        &       $5.91\pm1.01$       &   $6.90\pm0.93$   &      $6.50\pm0.97$      &         $6.97\pm0.54$         \\
             $\alphasyn$             &      $-0.97\pm0.21$       &  $-0.65\pm0.08$   &     $-0.89\pm0.19$      &         $-1.1\pm0.1$          \\
          EM (pc / cm$^6$)           &       $2.72\pm1.64$       &   $3.53\pm2.31$   &            -            &               -               \\
         $A_{\rm AME}$ [Jy]          &       $1.03\pm0.32$       &         -         &      $1.20\pm0.35$      &               -               \\
           $\nuame$ [GHz]            &      $17.15\pm3.17$       &         -         &     $17.65\pm3.30$      &               -               \\
               $\Wame$               &       $0.58\pm0.16$       &         -         &      $0.66\pm0.18$      &               -               \\
            $\tau_{353}$             &       $3.79\pm0.34$       &   $3.84\pm0.34$   &      $3.77\pm0.33$      &               -               \\
              $\betad$               &       $1.71\pm0.08$       &   $1.79\pm0.08$   &      $1.68\pm0.08$      &        $1.49\pm0.056$         \\
              $\Td$ [K]              &      $18.49\pm0.79$       &  $18.02\pm0.75$   &     $18.69\pm0.78$      &         $18.8\pm0.54$         \\ \hline
   $S_{\rm 1\,GHz}^{\rm ff}$ (Jy)    &       $0.39\pm0.24$       &   $0.51\pm0.33$   &            -            &         $0.33\pm0.26$         \\
    $S^{\rm AME}$ @ 25\,GHz (Jy)     &       $0.83\pm0.32$       &         -         &      $1.05\pm0.35$      &         $1.45\pm0.18$         \\

 $S_{\rm 3000\,GHz}^{\rm dust}$ (Jy) &       $3255\pm1130$       &   $3161\pm1098$   &      $3276\pm1103$      &         $3180\pm230$          \\
         $\chi^2_{\rm red}$          &           0.61            &       1.51        &          0.39           &             0.36              \\
\hline
\end{tabular}
    \label{table:SED_cases}
\end{table*}

\subsection{The global properties of M31 compared to the Milky Way}
\label{section:comparison_w_MW}
As pointed out in Section~\ref{section:introduction}, M31
is the galaxy most similar to the MW in the
Local Group. Therefore, it is straightforward to compare
the properties between the two, especially for AME,
which is the main focus of our study. First, we computed 
the AME emissivity for M31 as done in 
\cite{ameplanewidesurvey}, i.e. as the ratio
between the AME intensity at 28.4\GHz{} (in temperature
units) and the dust intensity at 100\mum{} from the fit: 
\begin{equation}
\emmAME= \frac{T_{\rm AME}^{\rm 28.4\,GHz}}{I_{100\,\mu\rm m}}=\frac{\frac{c^2}{2 k_{\rm B}\nu^2}S_{\rm AME}^{\rm 28.4\,GHz}}{S_{100\,\mu\rm m}}
\end{equation}
This ratio is intended to cancel the dependence on the 
column density, present in both AME and thermal dust
emission, thus allowing us to compare regions with 
extremely different morphologies. As pointed out by \cite{tibbs2013} this cancellation may not be perfect as this calculation of the AME emissivity is sensitive to the dust temperature. However thanks to the similarity between dust temperatures in M31 and in the MW this should not be a problem in our comparison. 

Differences between \citetalias{battistelli2019} and this work 
on dust parameters arise because of the different surveys used
(\textit{Herschel} and COBE-DIRBE, respectively). We find $\emmAME=
9.6\pm3.1\,\mathrm{\mu}$K/(MJy/sr), and compare it to
the results for the MW both from the \Planck{} collaboration
\citep{planck_GP_w_ancillary_data} and also Section~4.2 of 
\cite{ameplanewidesurvey}. \cite{planck_GP_w_ancillary_data}
quoted a $9.8\pm0.5\,\mathrm{\mu}$K/(MJy/sr) value\footnote{
There seems to be a typo on this \Planck{} paper, as in Section 7.3
the units for the emissivity are quoted as mK/(MJy/sr) instead of 
$\mathrm{\mu}$K/(MJy/sr).} for the 
MW, fully consistent with our M31 result, while \cite{
ameplanewidesurvey} returns $\emmAME=8.8\pm3.8\,\mathrm{\mu}$K/(MJy/sr),
which is a lower value, but still 
consistent because of the large uncertainties\footnote{
This estimate increases when only those pixels with 
strong AME detections are used, up to $\emmAME=11.6\pm3.5
\,\mathrm{\mu}$K/(MJy/sr)}. Both values are lower than
those previously found in several other works (see e.g. Table 3 of 
\citealt{amereview}), but the AME emissivity can show large 
variations (\citealt{davies2006} found variations as large as a 
factor 2, for example). Those variations would probably be embedded
within the integrated spectrum of the galaxy, effectively decreasing
the AME emissivity value with respect to that found in compact sources.

When computing the AME fraction at 28.4\GHz{} we find 
$S_{\rm AME}^{\rm 28.4\GHz{}}/S^{\rm 28.4\,GHz}_{\rm total
}=0.54\pm0.17$. This is slightly greater than the values 
obtained by \cite{ameplanewidesurvey} and 
\cite{planck_GP_w_ancillary_data}, $0.46\pm0.08$ and 
$0.45\pm0.01$ respectively, although it is consistent to within
$1\sigma$, owing  to the greater uncertainty in the M31 
estimate. Finally, when focusing on the synchrotron spectral
index, we find consistent estimates between M31 and the MW of
$-0.97_{-0.23}^{+0.18}$ and $-0.94\pm0.10$ 
\citep{ameplanewidesurvey} respectively.

\subsection{QUIJOTE-MFI M31 upper limits on polarization}
\begin{figure*}
    \hspace*{-0.75cm}
    \includegraphics[width=1.1\linewidth]{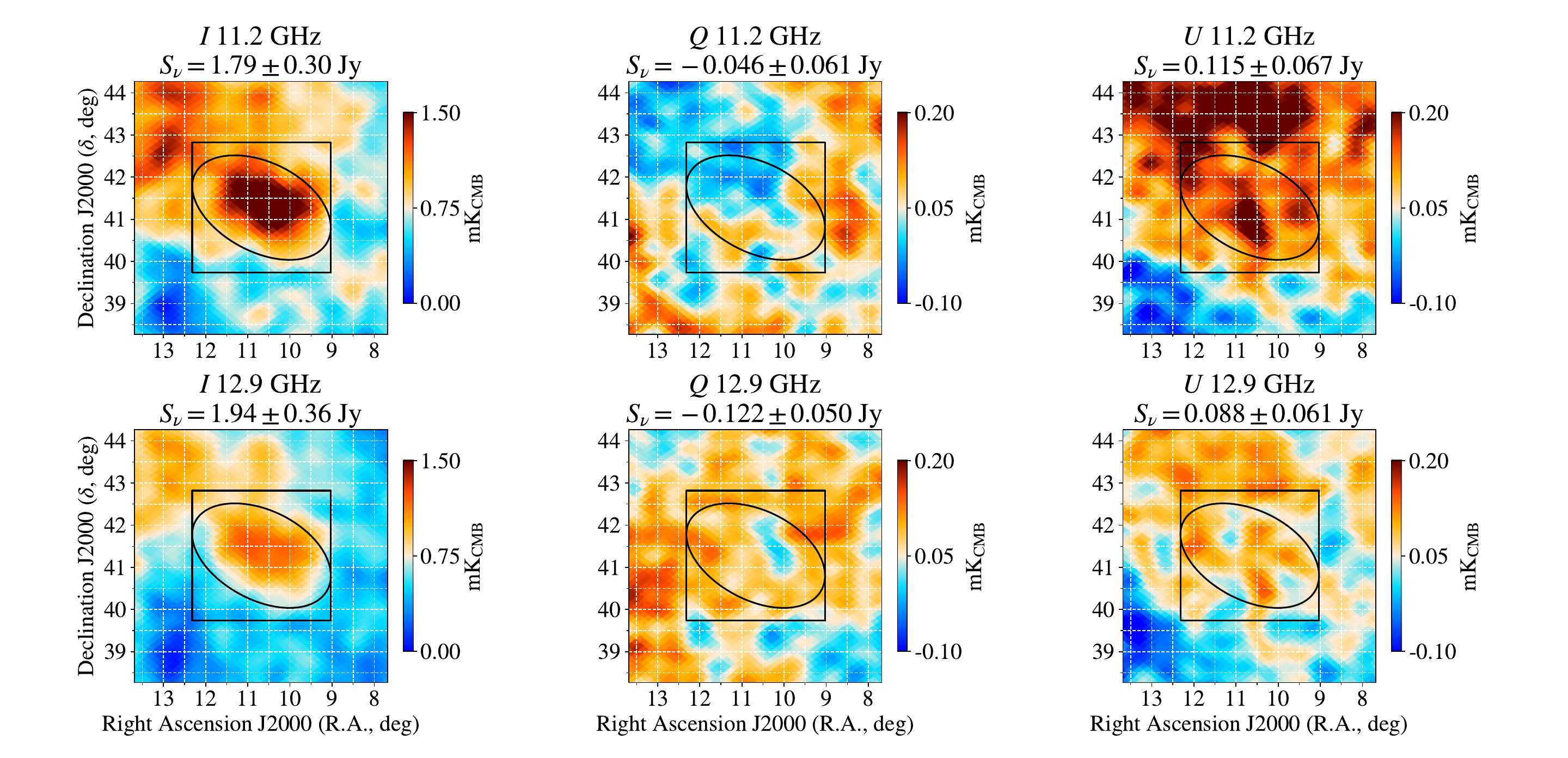}
    \caption{QUIJOTE-MFI maps at 11.2 and 12.9\GHz{} (top and 
    bottom rows, respectively) for the M31 field in intensity
    (left column) and Stokes parameters $Q$ (centre) and $U$ (right).
    The solid lines show the M31 (ellipse) and background 
    (rectangular) apertures defined in Section~\ref{section:AP}
    and used in Section~\ref{section:results}. These maps are 
    the raw ones after combining the data from the Wide Survey
    and raster observations: we have not subtracted neither the CMB 
    nor point sources.}
    \label{fig:IQU_maps}
\end{figure*}

As explained in Section~\ref{section:quijote-mfi}, the M31 field
has the best sensitivity values (shown in 
Table~\ref{tab:sensitivity}) of all the fields studied by
QUIJOTE-MFI so far, being slightly better than those from the Haze
\citep{hazewidesurvey} and W49, W51 and IC443 \citep{W51}
fields. This is because of the long integration times invested in those 
regions. Stokes $Q$ and $U$ maps of the M31 field are displayed in Fig.~\ref{fig:IQU_maps}.

Given the lack of any detectable 
polarized signal towards M31 on the QUIJOTE-MFI maps, we derived upper 
limits on its global polarized flux density. Using the same photometric 
approach and apertures as for 
intensity (Section~\ref{section:AP}), the measured Stokes $Q$ and 
$U$ parameters are $-0.046\pm0.061$\Jy{} and $0.115\pm0.067$\Jy{} at 
11\,GHz, and $-0.122\pm0.050$\Jy{} and $0.088\pm0.061$\Jy{} at 13\GHz{}. 
Except for the $Q$ value at 13\GHz{} showing a marginal detection, 
they are statistically consistent with zero. 
Taking the measured
values as upper limits, the noise debiased\footnote{Computed following 
\cite{Vaillancourt_poldebiased} and \cite{jarm_poldebiased}.} polarized
intensity estimates, $P$, are $\lt0.22$\Jy{} and $\lt0.18$\Jy{}, 
respectively at 11 and 13\GHz{} for a 95\% confidence limit (C.L.)). The flux density estimates at 
11 and 13\GHz{} are $1.79\pm0.30$\Jy{} and $1.94\pm0.36$\Jy{}
(after subtracting both the point sources and the CMB, as shown in
Table~\ref{table:maps}), 
so the polarization fractions for the integrated M31 flux density are 
below $12.5\%$ and $9.6\%$, respectively, for a 95\% C.L. Considering the AME contribution to the total flux in intensity,
the polarization fraction upper limits derived from this analysis 
are $\leq30\%$ at 11\GHz{} and $\leq20\%$ at 13\GHz{}. 
Although this upper limit is an order of magnitude 
greater than the most stringent ones obtained from the analysis of AME
sources in the MW \citep[e.g., ][]{W44, W51, perseus_raul}, this is the
first time that a limit on the AME polarization has been obtained for an
extragalactic object. It must be noted though that a potential underlying AME polarization in M31 in small angular scales may be largely smeared out in our aperture due to mixing of different polarization orientations. This would also be the case for a possible polarization of the synchrotron emission, that in specific regions of M31 is known to reach values of $\sim 60\%$ \citep{berkhuijsen2003}. In any case, owing to the sensitivity of our polarization data, and to the uncertainty in the determination of the integrated total-intensity flux density of the synchrotron emission, our derived constraints on the global synchrotron polarization fraction are above this level.

\section{Robustness tests}
\label{section:discussion}

\subsection{Impact of the selected CMB map on the AME amplitude}
\label{section:ame_dependence_on_cmb_map}
As first mentioned in Section~\ref{section:cmb_choice}, different
CMB maps yield different integrated measurements on the M31 aperture. 
These affect the photometry flux densities computed to build the SED as in 
Eq.~\ref{eq:AP}. We mentioned that the measurement from SMICA is the largest 
of the four maps, which implies a decrease in M31 flux densities at microwave
frequencies ($\nu\in(40,90)\GHz{}$).
This produces a steeper spectrum in this frequency range, and thus an AME component is
preferred over free-free or synchrotron. In the opposite scenario, the 
measurement from COMMANDER is lower, thus flattening the SED at those 
frequencies and decreasing AME amplitude.

We ran the same analysis as in Section~\ref{section:results} 
after using SMICA and COMMANDER instead of NILC to subtract the CMB 
anisotropies. AME amplitude increases to $\Aame=1.17\pm0.29\Jy{}$ 
when the SMICA map is used, and decreases to
$\Aame=0.92\pm0.34\Jy{}$ when using COMMANDER. Thus, the AME 
significance changes from $4.0\sigma$ to $2.7\sigma$ depending
on the choice of the CMB map. 
Besides, the higher $\Aame$ value when subtracting the signal
from SMICA implies a lower free-free amplitude and steeper synchrotron
index. In the COMMANDER case the situation is the opposite: the 
$\Aame$ decrease implies a larger free-free component and a flatter 
synchrotron spectrum. Because of the differences between the two cases,
we decided to use the NILC CMB map, which returns a $3.2\sigma$ $\Aame$
significance instead, as a trade-off scenario.

\subsection{QUIJOTE-MFI role in the fitting}
\begin{figure}
    \centering
    \includegraphics[width=1\linewidth]{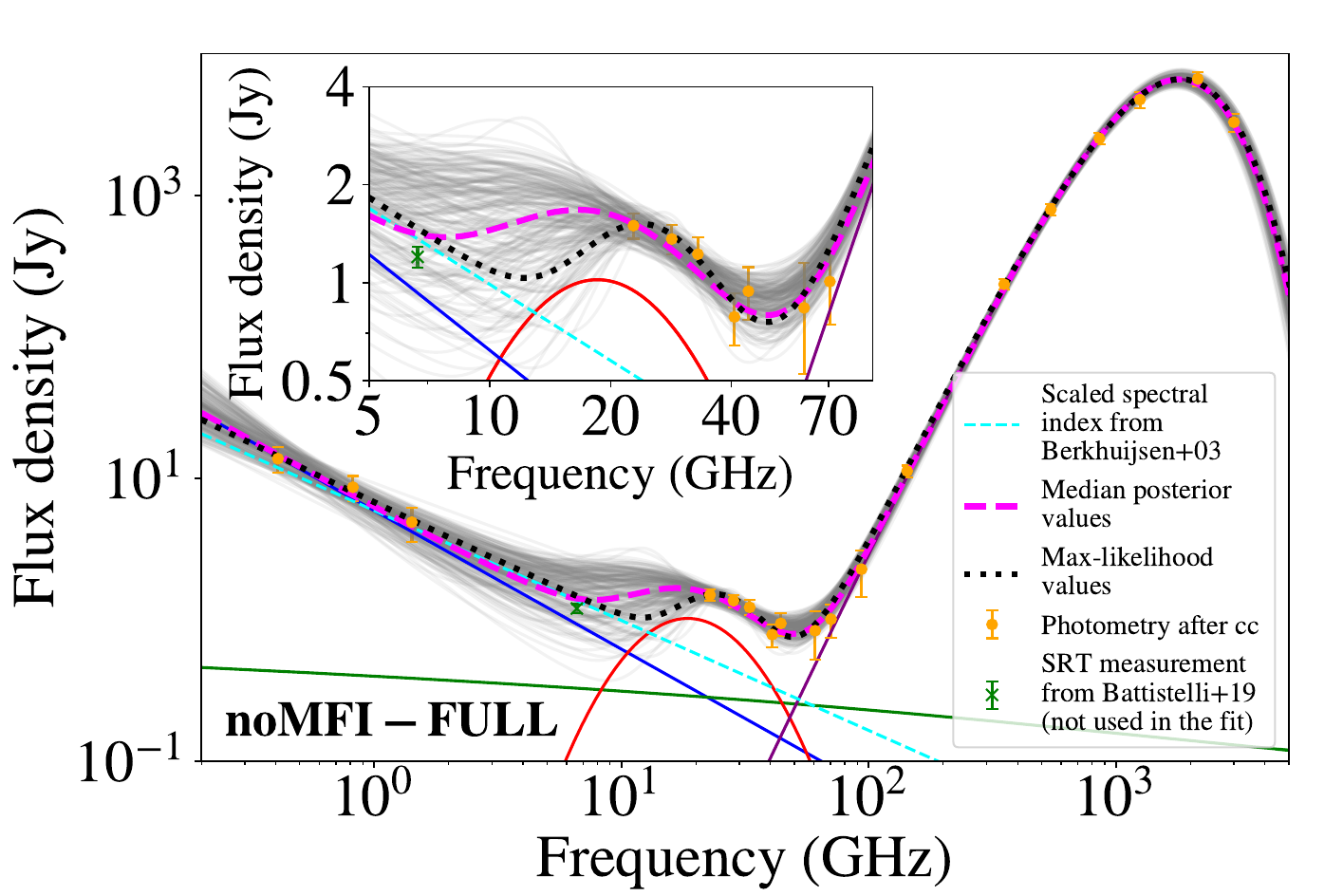}
    \includegraphics[width=1\linewidth]{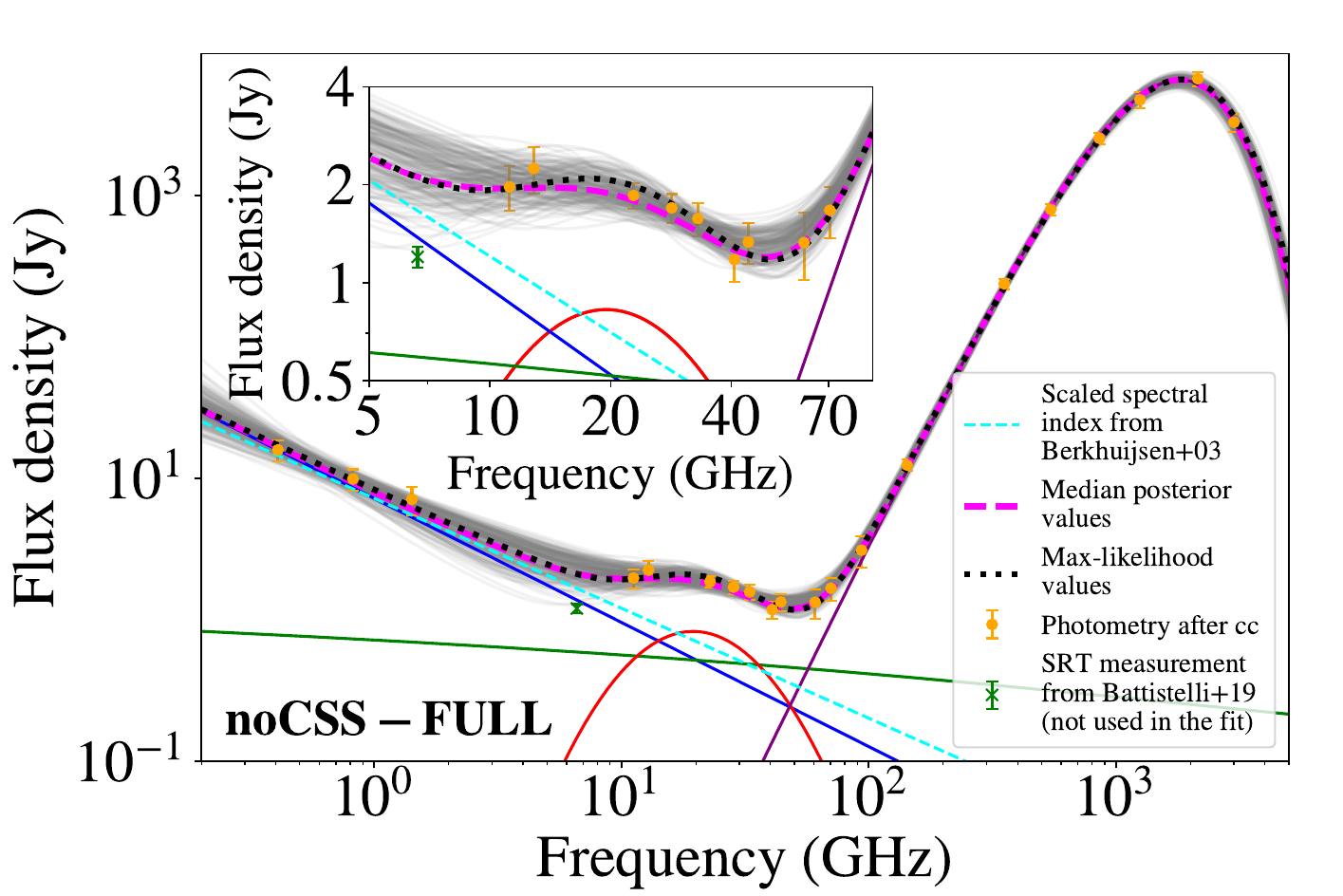}

    \caption{Top panel: same as 
    Fig.~\ref{fig:SED_cases_1}, but now without taking
    into account QUIJOTE-MFI data (noMFI). We can see how the
    dispersion of the model is now much greater between
    1 and 20\GHz{}, as expected. This causes a larger
    disagreement between the median and maximum likelihood fits
    than in the rest of the scenarios. Bottom panel: same,
    but now including QUIJOTE-MFI data and
    without subtracting point sources emission (noCSS).}
    \label{fig:SED_cases_noMFI}
\end{figure}

We repeated the former analyses without taking into 
account the data from QUIJOTE-MFI. In this way, we can 
assess its constraining power.
We show the results in Table~\ref{table:SED_cases_noMFI}
and the corresponding SED for the general (all emission
components considered) case in Fig.~\ref{fig:SED_cases_noMFI}.
The significance of the measurement of $\Aame$ is slightly 
reduced to $3.1\sigma$ when the QUIJOTE-MFI data are not used. 
The AME amplitude, $\Aame$, remains the same: it is now
$1.03\pm0.33$\Jy{} compared to the previous value of 
$1.03\pm0.32$\Jy{}. However, we find that the 
$\nuame$ posterior is now not constrained at lower values. This
implies an increase on its uncertainty, so 
$\nuame=18.3\pm5.8\GHz{}$, compared to $17.2\pm3.2\GHz{}$ 
when using QUIJOTE-MFI data. This is directly related to the absence of data in that frequency range (5--20\GHz{}). The larger uncertainty also implies larger differences between the preferred models when using the max-likelihood solution or that from the 1D distributions median values. 
The change is not as important for the 
AME width, $\Wame$, as it is $0.54\pm0.20$ now compared to the 
previous value of $0.58\pm0.16$.
Therefore, QUIJOTE-MFI data addition significantly reduces 
(almost by a factor 2) the uncertainty of the $\nuame$
parameter. Finally, we can see that the model 
providing the worst fit to the data is again the one trying
to account for all emission with just synchrotron and free--free
components and no AME component.

\begin{table*}
    \caption{Same as Table~\ref{table:SED_cases},
    but without taking into account QUIJOTE-MFI data.}
    \centering
    \begin{tabular}{ccccc}
\hline
                                     &  Case 1 (all components)  &  Case 2 (no AME)  &  Case 3 (no free-free)  &  From \cite{battistelli2019}  \\
\hline
        $S_{\rm 1\,GHz}$ (Jy)        &       $5.92\pm1.01$       &   $7.10\pm0.95$   &      $6.51\pm0.99$      &         $6.97\pm0.54$         \\
             $\alphasyn$             &      $-0.99\pm0.21$       &  $-0.65\pm0.08$   &     $-0.89\pm0.19$      &         $-1.1\pm0.1$          \\
          EM (pc / cm$^6$)           &       $2.77\pm1.65$       &   $3.19\pm2.15$   &            -            &               -               \\
         $A_{\rm AME}$ [Jy]          &       $1.03\pm0.33$       &         -         &      $1.20\pm0.35$      &               -               \\
           $\nuame$ [GHz]            &      $18.29\pm5.84$       &         -         &     $18.39\pm5.79$      &               -               \\
               $\Wame$               &       $0.54\pm0.20$       &         -         &      $0.63\pm0.22$      &               -               \\
            $\tau_{353}$             &       $3.79\pm0.34$       &   $3.85\pm0.35$   &      $3.76\pm0.33$      &               -               \\
              $\betad$               &       $1.71\pm0.08$       &   $1.79\pm0.08$   &      $1.68\pm0.08$      &        $1.49\pm0.056$         \\
              $\Td$ [K]              &      $18.49\pm0.78$       &  $18.01\pm0.75$   &     $18.72\pm0.79$      &         $18.8\pm0.54$         \\ \hline
   $S_{\rm 1\,GHz}^{\rm ff}$ (Jy)    &       $0.40\pm0.24$       &   $0.46\pm0.31$   &            -            &         $0.33\pm0.26$         \\
    $S^{\rm AME}$ @ 25\,GHz (Jy)     &       $0.87\pm0.42$       &         -         &      $1.06\pm0.41$      &         $1.45\pm0.18$         \\
 $S_{\rm 3000\,GHz}^{\rm dust}$ (Jy) &       $3219\pm1104$       &   $3155\pm1102$   &      $3284\pm1113$      &         $3180\pm230$          \\
              $\chi^2_{\rm red}$               &           0.62            &       1.53        &          0.39           &             0.36              \\
\hline
\end{tabular}
    \label{table:SED_cases_noMFI}
\end{table*}

\subsection{Source subtraction effect}
We repeated our analysis of the maps with no point source subtraction
applied, but keeping QUIJOTE-MFI data. This was done in order to
assess the impact of this subtraction on the level of AME. Results for this case are shown in 
Table~\ref{table:SED_cases_noCSS}. The general case (all emission
components considered) is also shown in the bottom panel of
Fig.~\ref{fig:SED_cases_noMFI}.
As it was the case when discarding QUIJOTE-MFI data, the uncertainties
on the AME parameters increased, thus reducing the 
significance for those parameters. There are slight changes in some parameters. In particular, the synchrotron becomes flatter, and as a result there is more synchrotron at microwave frequencies. The level of free-free is higher: as a consequence from both changes, the AME amplitude is lower. Therefore, the significance of $\Aame$ is
now lower too, down to $2.3\sigma$. The $\nuame$ and $\Wame$ values do not change 
significantly with respect to the main case presented in Section~\ref{section:results}: 
$19.5\pm4.3$\GHz{} and $0.58\pm0.26$, respectively, compared
to $17.2\pm3.2$\GHz{} and $0.58\pm0.16$. The changes of all parameters are always within $1\sigma$.


\begin{table*}
    \caption{Same as Table~\ref{table:SED_cases},
    but with no point source subtraction applied.}
    \centering
    \begin{tabular}{ccccc}
\hline
                                     &  Case 1 (all components)  &  Case 2 (no AME)  &  Case 3 (no free-free)  &  From \cite{battistelli2019}  \\
\hline
        $S_{\rm 1\,GHz}$ (Jy)        &       $7.21\pm1.15$       &   $7.62\pm1.13$   &      $8.17\pm0.98$      &         $6.97\pm0.54$         \\
             $\alphasyn$             &      $-0.88\pm0.18$       &  $-0.70\pm0.13$   &     $-0.76\pm0.16$      &         $-1.1\pm0.1$          \\
          EM (pc / cm$^6$)           &       $4.97\pm3.11$       &   $7.34\pm3.84$   &            -            &               -               \\
         $A_{\rm AME}$ [Jy]          &       $0.83\pm0.37$       &         -         &      $1.07\pm0.44$      &               -               \\
           $\nuame$ [GHz]            &      $19.49\pm4.26$       &         -         &     $20.58\pm4.51$      &               -               \\
               $\Wame$               &       $0.58\pm0.26$       &         -         &      $0.74\pm0.29$      &               -               \\
            $\tau_{353}$             &       $3.76\pm0.33$       &   $3.79\pm0.33$   &      $3.73\pm0.32$      &               -               \\
              $\betad$               &       $1.63\pm0.09$       &   $1.68\pm0.08$   &      $1.59\pm0.08$      &        $1.49\pm0.056$         \\
              $\Td$ [K]              &      $19.01\pm0.84$       &  $18.65\pm0.78$   &     $19.26\pm0.83$      &         $18.8\pm0.54$         \\ \hline
   $S_{\rm 1\,GHz}^{\rm ff}$ (Jy)    &       $0.72\pm0.45$       &   $1.06\pm0.56$   &            -            &         $0.33\pm0.26$         \\
    $S^{\rm AME}$ @ 25\,GHz (Jy)     &       $0.75\pm0.36$       &         -         &      $1.03\pm0.43$      &         $1.45\pm0.18$         \\

 $S_{\rm 3000\,GHz}^{\rm dust}$ (Jy) &       $3319\pm1165$       &   $3264\pm1108$   &      $3356\pm1129$      &         $3180\pm230$          \\
              $\chi^2_{\rm red}$               &           0.46            &       0.95        &          0.36           &             0.36              \\
\hline
\end{tabular}
    \label{table:SED_cases_noCSS}
\end{table*}

\subsection{Impact from changes in the background aperture}
\begin{figure}
    \centering
    \includegraphics[width=1\linewidth]{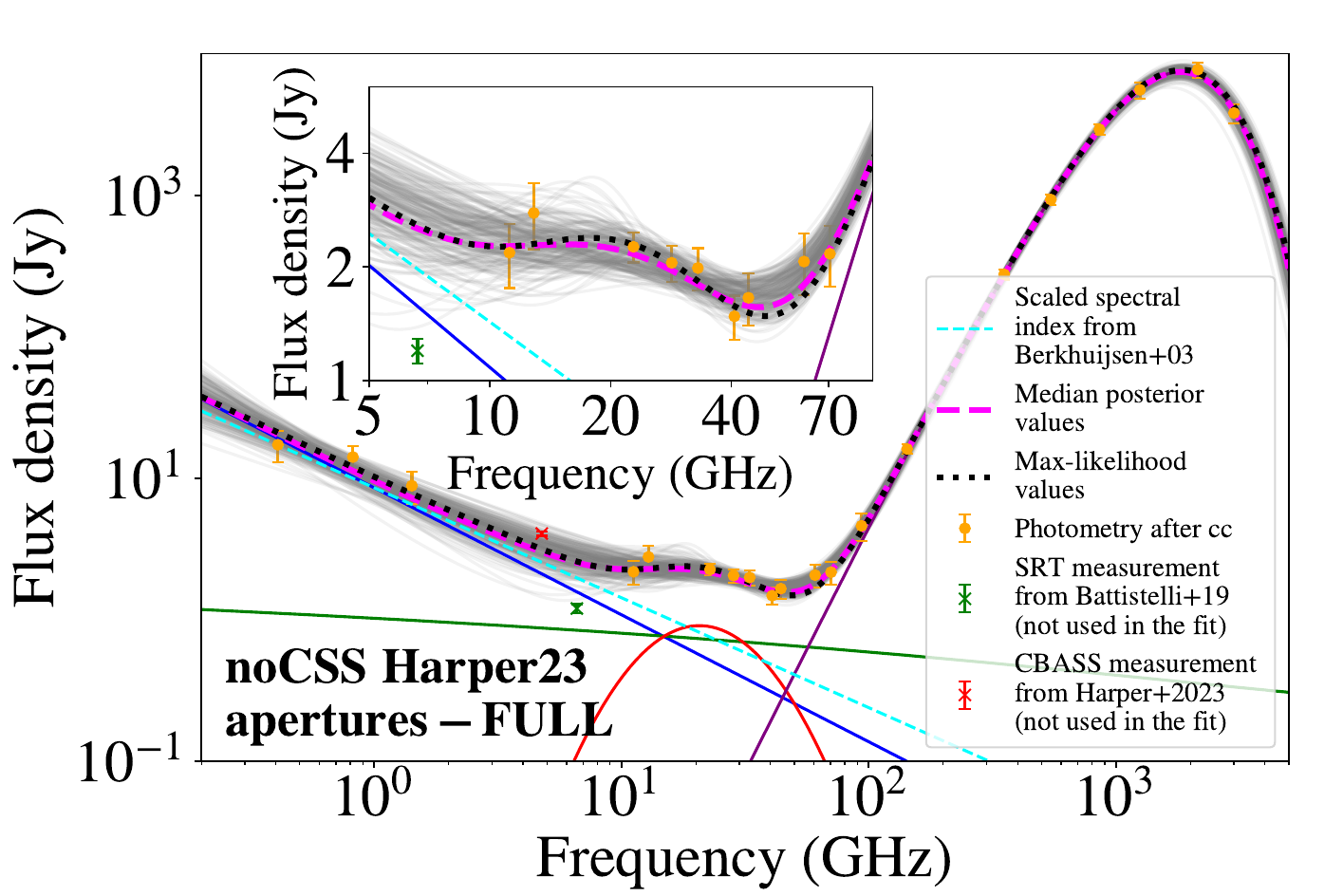}
    \includegraphics[width=1\linewidth]{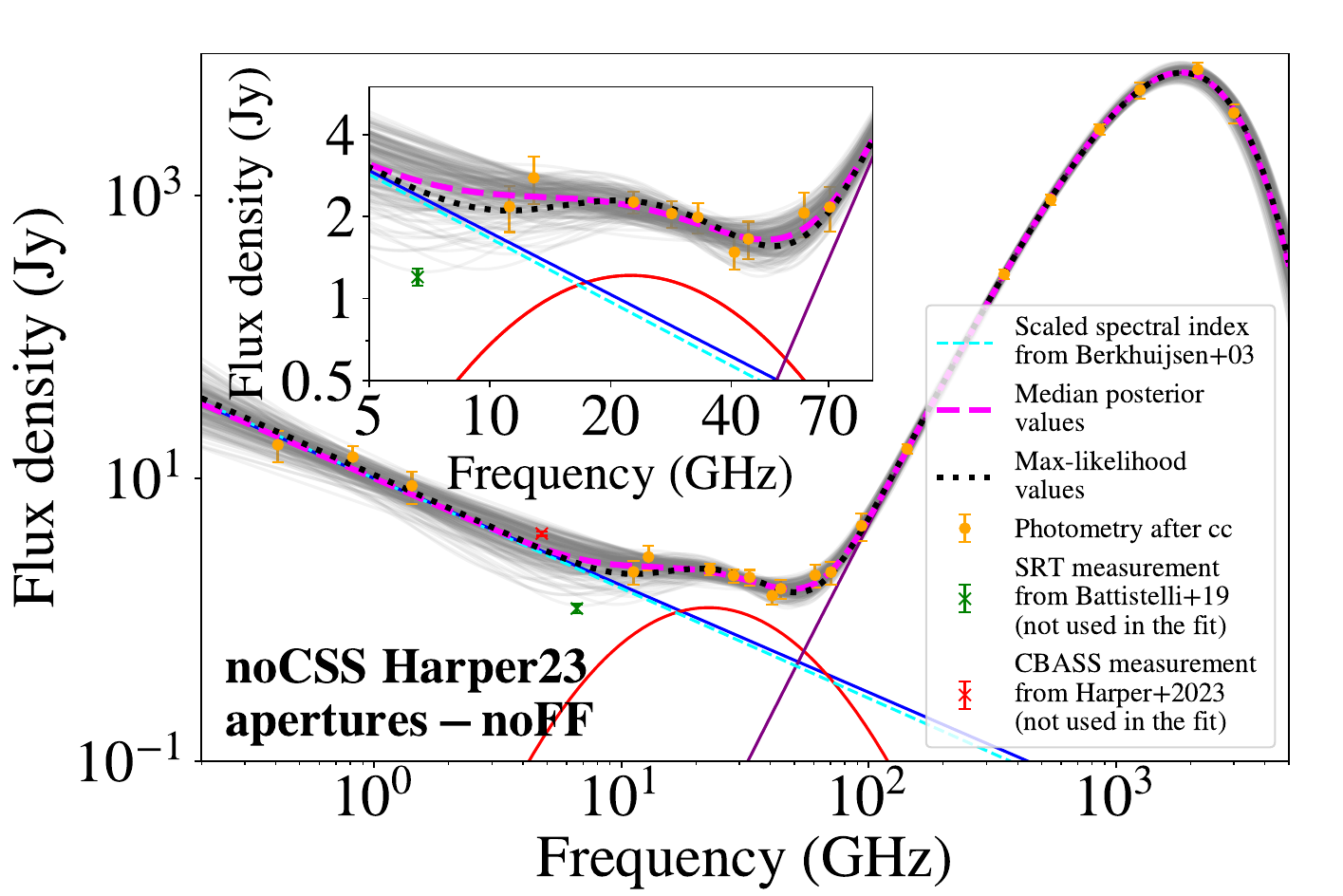}
    \includegraphics[width=1\linewidth]{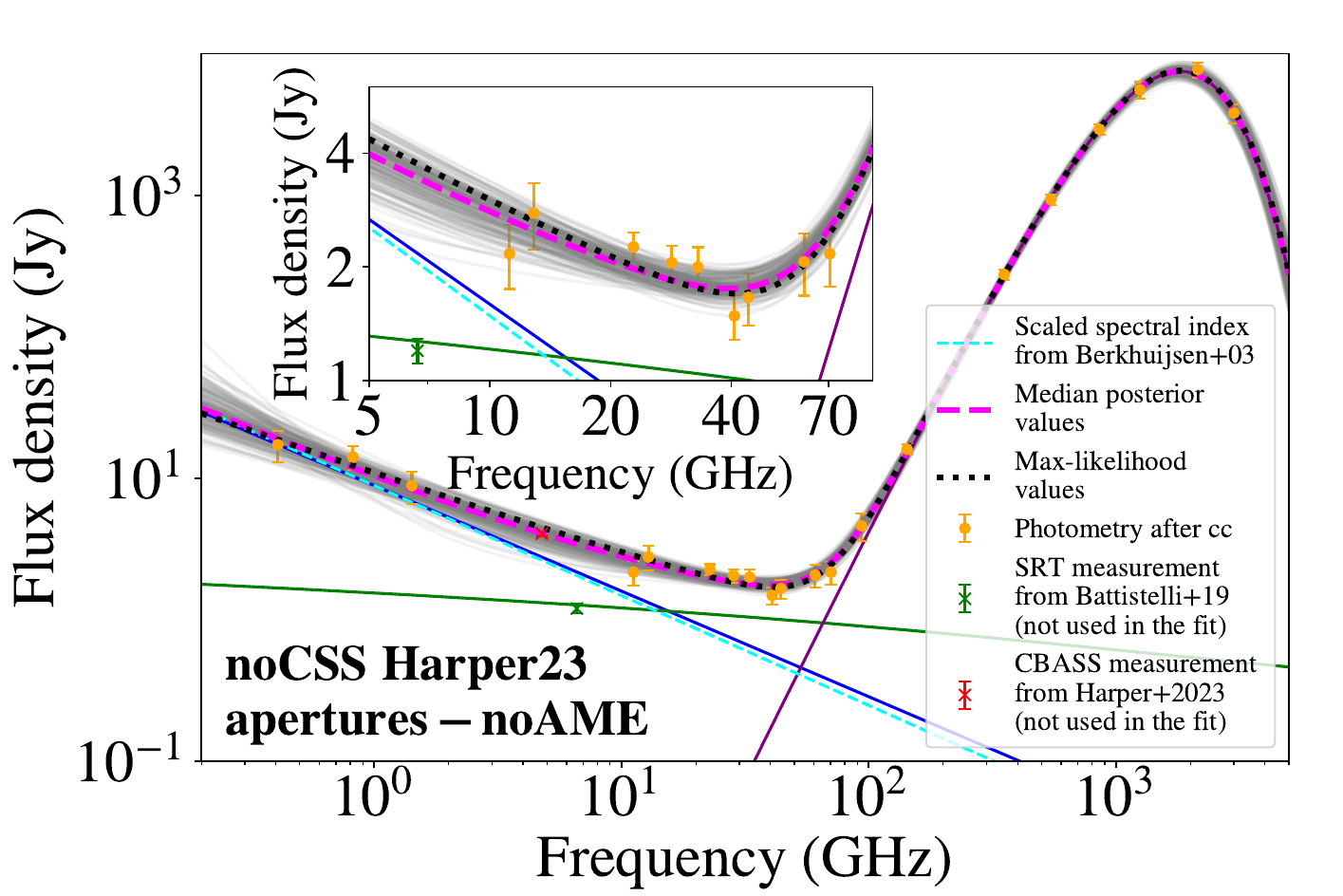}
    \caption{Same as Figures~\ref{fig:SED_cases_1} and
    \ref{fig:SED_cases_2}, 
    but now using an elliptical aperture far
    from M31 as the background region. Point
    source emission has not been subtracted for
    this case. This case is the one most similar to the analyses of
    \protect\cite{planckM31} and \protect\cite{cbassm31}.
    We are adding the C-BASS point from the
    latter just for comparison purposes.}
    \label{fig:SED_cases_noCSS_ell}
\end{figure}

Finally, we repeated the analysis using the same elliptical
aperture used in \cite{cbassm31}, which is also similar to
the one introduced in \cite{planckM31}. These regions have major 
semiaxes equal to 100, 110 and 154 arcminutes, the last 
two being  the inner and outer radii of the ring enclosing the background
region. The minor-to-major ratio is 0.7 for all three, with an 
east-to-north position angle equal to $-45\degr$. When comparing
with \cite{planckM31}, we are introducing the MFI data and 
subtracting the CMB using NILC as a template (plus adding an 
uncertainty term to account for the dispersion in the values of
the CMB templates, as explained in Section~\ref{section:cmb_choice}). 
Our main difference with \cite{cbassm31} is the addition of the
QUIJOTE-MFI raster data presented in Section~\ref{section:quijote-mfi},
while we lack the point from C-BASS at 4.76\GHz{}. That paper also 
used SMICA instead of NILC to account for CMB anisotropies: we
found that SMICA systematically returned higher estimates for the
AME than the rest of maps (following the discussions in Sections~\ref{section:cmb_choice}  and \ref{section:ame_dependence_on_cmb_map}). We are also not
keeping the point source subtraction because of the paucity
of objects in \cite{fatigoni2021} outside the SRT M31 field.
Results are shown in Table~\ref{table:SED_cases_noCSS_ell} (where
we show the case from \citealp{cbassm31} with no subtraction of
sources, as it is the most similar one to our analysis) and 
the corresponding SEDs in Fig.~\ref{fig:SED_cases_noCSS_ell}.
We recover an AME estimate with reduced significance ($2.2\sigma$)
only: $0.93\pm0.43$\Jy{}. Note though that the differences between our parameters and those of \citealp{cbassm31} are always below the $1\sigma$ level.

We find that the result in our study for the synchrotron index, 
$\alphasyn$, is consistent with those of \cite{berkhuijsen2003}
and \cite{planckM31} ($-0.90\pm0.20$ vs.\ $-1.0\pm0.2$ and 
$-0.92\pm0.16$ respectively), but steeper than that of
\cite{cbassm31} ($-0.63\pm0.05$). The synchrotron amplitude,
$\Asyn$ is consistent between this study and \cite{planckM31} and 
\cite{cbassm31}: $8.6\pm1.7$\Jy{} vs.\ $9.5\pm1.1$ and $9.7\pm0.6$, 
respectively. Our free--free estimate is larger, although consistent
with a non detection to within 2$\sigma$ too: $5.7\pm3.4$\,pc\,cm$^{-6}$
vs.\ $1.8\pm1.3$\,pc\,cm$^{-6}$ and $1.1\pm0.8$\,pc\,cm$^{-6}$. 
We get a similar consistency
in the AME amplitude with \cite{planckM31} ($S_{\rm AME}^{\rm 
30\GHz{}}=0.74\pm0.43\Jy{}$ vs.\ $S_{\rm AME}^{\rm 30\GHz{}}=
0.7\pm0.3$\,Jy), and the value from \cite{cbassm31} ($0.32\pm0.11$\,Jy)
is within $1\sigma$. The AME peak frequency value increased slightly 
from Section~\ref{section:results} to $\nuame=20.8\pm5.1$\GHz{}, 
while AME width decreased to $\Wame=0.55\pm0.29$. \cite{cbassm31}
found a higher value for $\nuame$, $25\pm2$\GHz{}. However, one must
take into account that our model includes an additional free
varying parameter driving the width of the AME distribution. On the
other hand, both \cite{planckM31} and \cite{cbassm31} fixed this 
value by assuming that the AME was fully explained by a single
template model (WIM and WNM templates from {\tt SPDUST} 
-\citealp{spdust1, spdust2}-, respectively). Our approach would be
more similar to a galaxy hosting multiple AME regions with different
spectral dependencies.

Finally, the AME
emissivity in this case can be also computed: we find $T_{\rm AME}^{\rm 
30\,GHz}/\taud=4.0\pm2.7$\,K/$\taud$, much higher than the one
obtained by \cite{cbassm31}. The $\emmAME$ estimate is lower than
that from Section~\ref{section:comparison_w_MW}, being 
$\emmAME=8.4\pm5.0$\,$\mu$K\,MJy$^{-1}$\,sr, although within
1$\sigma$, because of the large uncertainties.

When evaluating our fit at C-BASS nominal frequency, 4.76\,GHz, we find 
$S_{\rm 4.76\,GHz}=3.19_{-0.64}^{+0.64}$\,Jy, the large dispersion
being consistent with the lack of data between 2--10\,GHz. This difference 
with the value quoted in \cite{cbassm31} ($4.06\pm0.14$\,Jy) is between the 
differences for 1.42\,GHz and QUIJOTE-MFI surveys (1.5\,Jy and 0.55\,Jy 
respectively). 
It is noticeable in Fig.~\ref{fig:SED_cases_noCSS_ell} that the C-BASS measurement falls well within the uncertaintiy of our fitted model. Finally, it is worth noting that C-BASS and QUIJOTE are calibrated to different calibration scales. As explained in \cite{cbassm31} C-BASS data uses the model for Tau A of \cite{weiland2011}. On the contrary, as explained in \cite{mfiwidesurvey}, QUIJOTE data are calibrated to an updated model that is based on the same data used by \cite{weiland2011}, but including some improvements and the addition of Tau A flux densities from Planck data. Evaluation of the C-BASS model at 4.76\,GHz for epoch 2014.4 (C-BASS reference epoch, as observations took place between July 2013 and March 2015) yields a value that is $4.2\%$ higher than the prediction from the QUIJOTE model. Therefore rescaling the C-BASS point to the QUIJOTE model results in 3.89\,Jy, an even better consistency with the fitted models shown in Fig.~\ref{fig:SED_cases_noCSS_ell}.

\begin{table*}
    \caption{Same as Table~\ref{table:SED_cases},
    but using an elliptical aperture far from M31
    as the background region. Point source 
    subtraction was not applied in this case. 
    This is the scenario most similar to the analysis of
    \protect\cite{planckM31}. We show the results of
    \protect\cite{cbassm31} without applying the
    source subtraction, as it is the most similar
    case to ours.}
    \centering
    \begin{tabular}{cccccc}
\hline
                                     &  Case 1 (all components)  &  Case 2 (no AME)  &  Case 3 (no free-free)  &  \cite{planckM31}  &  \cite{cbassm31}  \\
\hline
        $S_{\rm 1\,GHz}$ (Jy)        &       $8.62\pm1.66$       &   $8.85\pm1.64$   &      $9.97\pm1.42$      &    $9.5\pm1.1$     &    $9.7\pm0.6$    \\
             $\alphasyn$             &      $-0.90\pm0.20$       &  $-0.75\pm0.17$   &     $-0.76\pm0.16$      &   $-0.92\pm0.16$   &  $-0.63\pm0.05$   \\
          EM (pc / cm$^6$)           &       $5.53\pm3.29$       &   $8.64\pm3.88$   &            -            &    $1.8\pm1.3$     &    $1.1\pm0.8$    \\
         $A_{\rm AME}$ [Jy]          &       $0.92\pm0.43$       &         -         &      $1.22\pm0.52$      &         -          &         -         \\
           $\nuame$ [GHz]            &      $20.84\pm5.06$       &         -         &     $22.46\pm5.03$      &         -          &     $25\pm2$      \\
            $\tau_{353}$             &       $3.50\pm0.32$       &   $3.53\pm0.31$   &      $3.46\pm0.30$      &         -          &         -         \\
              $\betad$               &       $1.53\pm0.08$       &   $1.58\pm0.08$   &      $1.50\pm0.08$      &   $1.62\pm0.11$    &   $1.76\pm0.05$   \\
              $\Td$ [K]              &      $19.59\pm0.88$       &  $19.27\pm0.82$   &     $19.86\pm0.87$      &    $18.2\pm1.0$    &   $17.3\pm0.4$    \\ \hline
   $S_{\rm 1\,GHz}^{\rm ff}$ (Jy)    &       $1.00\pm0.59$       &   $1.56\pm0.70$   &            -            &         -          &         -         \\
    $S^{\rm AME}$ @ 25\,GHz (Jy)     &       $0.74\pm0.43$       &         -         &      $1.13\pm0.52$      &    $0.7\pm0.3$     &   $0.32\pm0.11$   \\
 $S_{\rm 3000\,GHz}^{\rm dust}$ (Jy) &       $3933\pm1358$       &   $3874\pm1295$   &      $3977\pm1322$      &    $3340\pm440$    &         -         \\
              $\chi^2_{\rm red}$               &           0.77            &       1.12        &          0.71           &        1.1         &        2.6        \\
\hline
\end{tabular}
    \label{table:SED_cases_noCSS_ell}
\end{table*}

\section{Conclusions}
\label{section:conclusions}
In this study we used new data from the QUIJOTE-MFI experiment
at 10--20\GHz{} to study the Andromeda Galaxy (M31). We
built the integrated SED in intensity of M31 between 0.408 and 
3000\GHz{},
and performed a component separation analysis after subtracting
the emission from CMB anisotropies and point sources
in the region present in the \cite{fatigoni2021} catalogue. We measure
the AME amplitude with a significance of $3.2\sigma$ in the 
integrated SED of M31, independently of the previous 
\citetalias{battistelli2019} detection. 
Moreover, our fitted model is consistent with
the 6.6\GHz{} measurement provided by that study.
The addition of QUIJOTE-MFI data improves the definition
of AME spectral parameters: its peak frequency, 
$\nuame=17.2\pm3.2$\GHz{}, and width, 
$\Wame=0.58\pm0.16$. $\nuame$ is low for most theoretical
models but is consistent with most recent studies pointing to
$\nuame$ being below 21\GHz{} \citep[e.g.][]{cbassSH2022, 
ameplanewidesurvey}. However, $\Wame$ is slightly 
greater than expected from theoretical models, but again
consistent with recent studies \citep[e.g.][]{AMEwidesurvey, 
ameplanewidesurvey}. This can be explained by the probable 
large number of AME regions within M31 with different $\nuame$
and $\Wame$ values, which naturally broaden the distribution
of the integrated measurement. Nevertheless, the significance of the AME
detection is not as dependent on the QUIJOTE-MFI data addition
as on the subtraction of point sources. The significance of the
measurement of $\Aame$ is still greater than 3$\sigma$ ($3.1\sigma$) 
when discarding the QUIJOTE-MFI data, but this significance falls
below 3$\sigma$ ($2.3\sigma$) when no point source subtraction
is applied. The $\Aame$ significance also falls below $3\sigma$
when the COMMANDER map is used to subtract the CMB anisotropies instead of NILC or SMICA, highlighting
the importance of the choice of the CMB map. When studying the AME
emissivity in M31, we found $\emmAME=9.6\pm3.1\,\mathrm{\mu}$K/(MJy/sr)
compared to the $\emmAME=8.8\pm3.8\,\mathrm{\mu}$K/(MJy/sr) estimate 
obtained by \cite{ameplanewidesurvey} for the Milky Way, 
pointing to a similar AME behaviour in both. 

We also provided a 
similar comparison to \cite{cbassm31}, who recently measured
AME in M31 again, although with much lower amplitude than the previous studies of
\citetalias{battistelli2019} and of \cite{planckM31}. 
This difference is driven mainly by the lower flux density derived from SRT data in \citetalias{battistelli2019} as compared with the C-BASS value of \cite{cbassm31}. However, the two works also show important differences in their data analysis processes. Especially, the different apertures used to perform aperture photometry are of extremely importance. We find consistent results with both when replicating each of the methodologies, while being independent from them - i.e. not using their data. 
The AME emissivity estimate from \cite{cbassm31} is off by a factor larger than 20 when
compared with previous measurements \citep{planck_GP_w_ancillary_data, 
cbassSH2022, ameplanewidesurvey}. When replicating the conditions from
\cite{cbassm31}, we obtain an AME emissivity estimate
$\emmAME=8.4\pm5.0\,\mathrm{\mu}$K/(MJy/sr), lower than the previous value 
of $\emmAME=9.6\pm3.1\,\mathrm{\mu}$K/(MJy/sr)
probably due to the larger importance of the outskirts 
of M31. This probably explains why this value is also lower than the 
estimate for the MW: when observing the MW from its inside, we are not
sensitive to these outer regions. This emissivity estimate is equal to
$4.0\pm2.7$\,K/$\taud$ in
$T_{\rm AME}^{\rm 30\,GHz}$ units, more in line with previous measurements 
\citep{davies2006, planck_GP_w_ancillary_data, 
planckemissivityestimates, hensley2017, cbassSH2022, ameplanewidesurvey}
than the one from \cite{cbassm31}.

M31 remains the only detection to date of AME in the integrated spectrum of an
external galaxy. We have also provided upper limits for the polarization of
this AME component. Although these limits are large when compared with
those from brighter AME sources (mostly within our Galaxy), this is the
first time that such upper limits have been computed for an extragalactic
object. The results of this work 
encourage the reproduction of this kind of of measurement for
galaxies with different sets of properties throughout the lifetime of the Universe.

\section*{Acknowledgements}
We thank the staff of the Teide Observatory for invaluable assistance in the commissioning and operation of QUIJOTE.
The {\it QUIJOTE} experiment is being developed by the Instituto de Astrofisica de Canarias (IAC),
the Instituto de Fisica de Cantabria (IFCA), and the Universities of Cantabria, Manchester and Cambridge.
Partial financial support was provided by the Spanish Ministry of Science and Innovation 
under the projects AYA2007-68058-C03-01, AYA2007-68058-C03-02, AYA2010-21766-C03-01, AYA2010-21766-C03-02,
AYA2014-60438-P,  ESP2015-70646-C2-1-R, AYA2017-84185-P, ESP2017-83921-C2-1-R, PID2019-110610RB-C21, PID2020-120514GB-I00,
IACA13-3E-2336, IACA15-BE-3707, EQC2018-004918-P, the Severo Ochoa Programs SEV-2015-0548 and CEX2019-000920-S, the
Maria de Maeztu Program MDM-2017-0765 and by the Consolider-Ingenio project CSD2010-00064 (EPI: Exploring
the Physics of Inflation).
We acknowledge support from the ACIISI, Consejeria de Economia, Conocimiento y Empleo del Gobierno de Canarias and the European Regional Development Fund (ERDF) under grant with reference ProID2020010108.
This project has received funding from the European Union's Horizon 2020 research and innovation program under
grant agreement number 687312 (RADIOFOREGROUNDS).

We thank the anonymous referee whose comments helped 
to improve this work. MFT acknowledges support from from the 
Agencia Estatal de Investigación (AEI) of the Ministerio de 
Ciencia, Innovación y Universidades (MCIU) and the European 
Social Fund (ESF) under grant with reference PRE-C-2018-0067.
C.A-T acknowledges support from the Millennium Nucleus on 
Young Exoplanets and their Moons (YEMS).
FP acknowledges support from the Agencia Canaria de Investigación, 
Innovación y Sociedad de la Información (ACIISI) under the European 
FEDER (FONDO EUROPEO DE DESARROLLO REGIONAL) de Canarias 2014-2020 
grant No. PROID2021010078.
We acknowledge the use of data provided by the Centre d'Analyse
de Données Etendues (CADE), a service of IRAP-UPS/CNRS 
(http://cade.irap.omp.eu,~\citealt{CADE}).
This research has made use of the SIMBAD database,
 operated at CDS, Strasbourg, France~\citep{simbad}. 
Some of the results in this paper have been derived using 
the healpy and {\tt HEALPix} packages~\citep{Healpix, 
Healpix2}. We have also used {\tt scipy}~\citep{scipy}, 
{\tt emcee}~\citep{emcee}, {\tt numpy}~\citep{numpy},
 {\tt matplotlib}~\citep{matplotlib}, {\tt corner}~\citep{corner} 
and {\tt astropy}~\citep{astropy1, astropy2} \textsc{python}
packages.

\section*{Data availability}
The QUIJOTE raster scan data used in this paper are property of 
the QUIJOTE Collaboration and can only be shared on request to 
the corresponding authors. All M31 fits are available upon reques
to the QUIJOTE collaboration. QUIJOTE-MFI Wide Survey data is 
available in the QUIJOTE collaboration webpage 
({\tt https://research.iac.es/proyecto/quijote}).

\bibliographystyle{mnras}
\bibliography{quijote,m31_biblio}

\bsp	
\label{lastpage}
\end{document}